\documentclass[12pt,notitlepage,aps,amsmath,amssymb]{revtex4-1} 
\usepackage{graphicx}				
\usepackage{caption,subcaption}

\usepackage{natbib} 

\newcommand{\Sn}{\Phi}
\newcommand{\sn}{\phi} 
\newcommand{\nSn}{ {\widetilde{\Phi}} }
\newcommand{\rSn}{ \Sn_\text{rms}}
\newcommand{\Snn}{\Psi}
\newcommand{\snn}{\psi}
\newcommand{\nSnn}{ {\widetilde{\Snn}} }
\newcommand{\nsnn}{ {\widetilde{\snn}} }
\newcommand{\rSnn}{ \Snn_\text{rms}}

\newcommand{\mA}{\left< A \right>}
\newcommand{\m}[1]{\left< #1 \right>}
\newcommand{\tm}[1]{\langle #1 \rangle}
\newcommand{\hyp}{ M }

\newcommand{\td}{\tau_\text{d}}
\newcommand{\tw}{\tau_\text{w}}
\newcommand{\dt}{\triangle_\text{t}}

\newcommand{\abs}[1]{\lvert #1 \rvert}
\newcommand{\D}{\text{d}}
\newcommand{\Lin}{\mathcal{L}}
\newcommand{\four}[1]{\mathcal{F}\left[{#1}\right](\omega)}
\newcommand{\fourinv}[1]{\mathcal{F}^{-1}\left[{#1}\right](\tau)}
\newcommand{\fourT}[1]{\mathcal{F}_T\left[{#1}\right](\omega)}
\newcommand{\psd}[1]{\mathcal{S}_{#1}(\omega)}
\newcommand{\obs}{\Omega}
\newcommand{\dyn}{\Delta}
\newcommand{\rms}{\text{rms}}
\newcommand{\norm}[1]{ {\widetilde{#1}} }

\let\originalleft\left
\let\originalright\right
\renewcommand{\left}{\mathopen{}\mathclose\bgroup\originalleft}
\renewcommand{\right}{\aftergroup\egroup\originalright}

\newcommand{\Ref}[1]{Ref.~\onlinecite{#1}}
\newcommand{\Refs}[1]{Refs.~\onlinecite{#1}}
\newcommand{\Eqref}[1]{Eq.~\eqref{#1}}
\newcommand{\Eqsref}[1]{Eqs.~\eqref{#1}}
\newcommand{\Figref}[1]{Fig.~\ref{#1}}
\newcommand{\Figsref}[1]{Figs.~\ref{#1}}
\newcommand{\Secref}[1]{Sec.~\ref{#1}}
\newcommand{\Appref}[1]{Appendix~\ref{#1}}

\begin{document}

\title{Statistical properties of a filtered Poisson process with additive random noise: Distributions, correlations and moment estimation}

\author{A.~Theodorsen}
\email{audun.theodorsen@uit.no}
\affiliation{Department of Physics and Technology, UiT The Arctic University of Norway, N-9037 Troms{\o}, Norway}

\author{O.~E.~Garcia}
\email{odd.erik.garcia@uit.no}
\affiliation{Department of Physics and Technology, UiT The Arctic University of Norway, N-9037 Troms{\o}, Norway}

\author{M.~Rypdal}
\email{martin.rypdal@uit.no}
\affiliation{Department of Mathematics and Statistics, UiT The Arctic University of Norway, N-9037 Troms{\o}, Norway}

\date{\today}

\begin{abstract}
  Filtered Poisson processes are often used as reference models for intermittent fluctuations in physical systems. Such a process is here extended by adding a noise term, either as a purely additive term to the process or as a dynamical term in a stochastic differential equation. The lowest order moments, probability density function, auto-correlation function and power spectral density are derived and used to identify and compare the effects of the two different noise terms. Monte-Carlo studies of synthetic time series are used to investigate the accuracy of model parameter estimation and to identify methods for distinguishing the noise types. It is shown that the probability density function and the three lowest order moments provide accurate estimations of the parameters, but are unable to separate the noise types. The auto-correlation function and the power spectral density also provide methods for estimating the model parameters, as well as being capable of identifying the noise type. The number of times the signal crosses a prescribed threshold level in the positive direction also promises to be able to differentiate the noise type.
\end{abstract}

\maketitle

\section{Introduction}
Intermittent fluctuations are found in a variety of physical systems such as atmospheric winds \cite{narasimha-07}, astrophysical plasmas \cite{robinson-95, kristensen-91}, fission chambers \cite{elter-14}, diodes and electric circuits \cite{rice-44,rice-45} and magnetic confinement experiments \cite{endler-jnm-95,carreras-05,liewer-85,dippolito-11,zweben-07}, as well as in fields such as finance \cite{jang} and physiology \cite{fesce}. In several such systems, treatments of intermittent effects as a superposition of random variables has been fruitful, see for instance \Refs{hole-04,hole-06}. In this contribution, we will focus on a particular reference model for intermittent fluctuations, the filtered Poisson process (FPP) (also called a generalized shot noise process) \cite{elter-14,rice-44,jang,fesce}. This model consists of a super-position of uncorrelated pulses with a uniform pulse shape and randomly distributed pulse amplitudes, arriving  according to a Poisson process \cite{garcia-prl-12,garcia-pop-16}. The FPP has been considered by e.~g.~\Refs{bondesson-82,pecseli-fps,parzen,kotler-99,lowen-fbpp}.

This contribution is primarily motivated by turbulent flows in the far scrape-off layer of magnetically confined plasmas. Evidence points towards these fluctuations being caused by filamentary structures moving radially outwards, transporting particles and heat through the scrape-off layer towards main chamber walls \cite{carreras-05,dippolito-11,zweben-07}. Time series obtained from probe measurements and gas puff imaging diagnostics exhibit similar behavior for a wide range of machine parameters, having skewed and flattened probability distribution functions (PDFs) resembling Gamma distributions and large amplitude fluctuations with exponential pulse shapes and exponentially distributed amplitudes, arriving according to a Poisson process \cite{garcia-pop-13,garcia-jnm-13,graves-05,kube-16,theodorsen-ppcf-16,garcia-nf-07}. 
The FPP with exponentially distributed pulse amplitudes and a pulse shape consisting of a rapid rise and exponential decay can be shown to be Gamma distributed \cite{garcia-prl-12,garcia-pop-16,bondesson-82}. By adding an independent, normally distributed variable to the process, the PDF of the resulting process is a convolution of a Gamma PDF and a normal PDF. This result has been shown to be in very good agreement with probe measurements from the Alcator C-Mod and KSTAR tokamaks \cite{kube-16,garcia-kstar}.

In this contribution, we will extend the reference FPP model by adding normally distributed noise in two different ways, either as a purely additive term to the process, modeling measurement noise or other processes unconnected to the dynamics of the FPP, or as a dynamical noise term in the stochastic differential equation for the reference model, resembling an Ornstein-Uhlenbeck process. We will mainly consider how these different noise terms affect the PDF of the resulting signal, its auto-correlation function and its power spectral density. Additionally, we will consider the rate at which the processes cross a certain threshold level in the positive direction. The goal of this contribution is to find methods for discriminating the two types of noise and to identify reliable methods for estimating the model parameters in a given realization of the process.

This contribution is organized as follows: In \Secref{sec:fpp}, the pure FPP is considered. In \Secref{sec:noise}, the two types of noise are considered. The lowest order moments and the PDF of the FPP with additional noise are discussed in \Secref{sec:momdist}, and the power spectral density and auto-correlation function of this process are discussed in \Secref{sec:psd}. In order to differentiate the types of noise and to compare different methods of parameter estimation, Monte-Carlo studies of synthetic data are presented in \Secref{sec:synth}. \Secref{sec:conclusion} concludes the contribution. In \Appref{app:results}, a list of symbols and the most important analytical results of this contribution are collected. \Appref{app:psd} contains derivations relating to the power spectral densities and auto-correlation functions discussed.

\section{Filtered Poisson process}\label{sec:fpp}
In this section, we present the FPP to be analyzed in this contribution. This process is constructed as a super-position of $K$ pulses arriving in a time interval $[0,T]$:
\begin{equation}
  \Sn_K(t) = \sum_{k=1}^{K(T)} A_k \varphi\left( \frac{t-t_k}{\td} \right).
  \label{eq:shot_noise}
\end{equation}
where the pulse duration time $\td$ is taken to be the same for all pulses. The pulse amplitudes $A_k$ are taken to be exponentially distributed with mean value $\tm{A}$,
\begin{equation}
  P_A(A;\mA) = \frac{1}{\mA}\exp\left( - \frac{A}{\mA} \right),
  \label{eq:amp_exp}
\end{equation}
where $A \geq 0$ and $\tm{\bullet}$ here and in the following indicates the average over all random variables. 

As an idealization of a pulse with a fast rise and an exponential decay, we use the one-sided pulse form
\begin{equation}
  \varphi\left( \eta \right) =\Theta(\eta) \exp\left(- \eta \right),
  \label{eq:pulse_shape}
\end{equation}
where $\Theta$ is the Heaviside step function and $\eta$ is a dimensionless variable. Compared to a two-sided exponential pulse function, this simplification does not affect the moments or PDF of the process, simplifies the auto-correlation and power spectra (see \Ref{theodorsen-ppcf-16} for the auto-correlation of  this process with finite growth) and allows the formulation of the process by a simple stochastic differential equation as described below.

 The pulses are assumed to arrive according to a Poisson process with constant rate ( see for example Refs.~\onlinecite{pecseli-fps},~\onlinecite[Ch.~4.1]{parzen}~or~\onlinecite[p.~562]{stark-prob}). Thus the (non-negative) number of arrivals $K(T)$ is Poisson distributed,
\begin{equation}
  P_K(K;T,\tw) = \frac{1}{K!} \left( \frac{T}{\tw} \right)^K \exp\left( -\frac{T}{\tw} \right),
  \label{eq:poisson dist}
\end{equation}
where we have taken the mean value to be $\tm{K}=T/\tw$. It can be shown that the waiting time between consecutive pulses is exponentially distributed with mean value $\tw$ \cite[p.~135]{parzen}, while the $K$ arrival times $t_k$ are independent and uniformly distributed on the interval $\left[ 0,T \right]$ \cite[p.~140]{parzen}. The ratio between pulse duration time and average time between pulses,
\begin{equation}
  \gamma = \frac{\td}{\tw},
  \label{eq:intermittency-parameter}
\end{equation}
is in the following referred to as the \emph{intermittency parameter}.

\begin{figure}
  \centering
  \includegraphics[width = 0.6\textwidth]{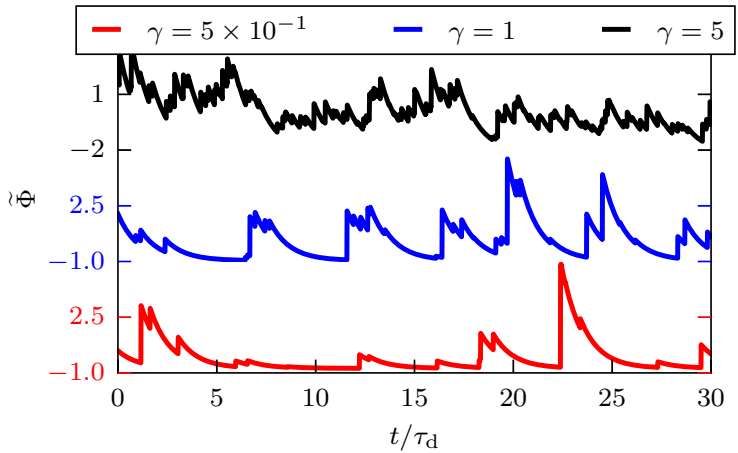}
  \caption{\label{fig:S_e0_dt0.01_mult_gamma}Synthetically generated pure filtered poisson processes with $\theta = 10^{-2}$ and various values of $\gamma$.}
\end{figure}

While the FPP is a continuous process, any experimental data or synthetic realization of the process is discrete. As the time resolution of numerical data may be important for the noise processes, we also introduce the normalized time step,
\begin{equation}
  \theta = \frac{\dt}{\td},
  \label{eq:norm-time-step}
\end{equation}
where $\dt$ is the time step for synthetically generated signals. Details on the synthetically generated signals will be discussed in \Secref{sec:synth}. Some realizations of $\Sn_K(t)$ for various values of $\gamma$ are presented in \Figref{fig:S_e0_dt0.01_mult_gamma}, where we have used the normalization
\begin{equation}
  \nSn = \frac{\Sn-\m{\Sn}}{\rSn}.
  \label{eq:resc_sgnl}
\end{equation}
Here and in the following, we will use a tilde to denote a normalized variable with zero mean and unit standard deviation.
For $\gamma < 1$, pulses arrive rarely and the signal spends large amounts of time close to zero value, resulting in a strongly intermittent signal. For $\gamma > 1$, pulses overlap and the signal begins to resemble random and symmetric fluctuations around a mean value.

It can be shown that the stationary PDF of the random variable $\Sn$ is a gamma distribution \cite{garcia-prl-12,garcia-pop-16};
\begin{equation}
  P_\Sn(\sn;\gamma,\mA) = \frac{1}{\mA \Gamma(\gamma)}\left( \frac{\sn}{\mA} \right)^{\gamma-1} \exp\left( -\frac{\sn}{\mA} \right),
  \label{eq:sn_pdf}
\end{equation}
where $\sn > 0$ due to the non-negative pulse amplitudes and pulse functions. The lowest order moments of $\Sn$ are the mean $\tm{\Sn} = \gamma \mA$, the standard deviation $\rSn = \gamma^{1/2} \mA$, the skewness $S_\Sn = 2/\gamma^{1/2}$ and the flatness $F_\Sn = 3+6/\gamma$, giving a parabolic relationship between the skewness and flatness moments of $\Sn$ \cite{garcia-prl-12,garcia-pop-16},
\begin{equation}
	F_\Sn = 3+3 S_\Sn /2.
	\label{eq:parabolic_fpp}
\end{equation}
In \Secref{sec:momdist} it will be shown how additive noise alters this relationship.

It is possible to write the FPP as a stochastic differential equation. In the case of a one-sided exponential pulse function, it takes the form \cite{kontorovich-96}
\begin{equation}
 \td\, \frac{\D\Sn_K}{\D t} = - \Sn_K + \sum_{k = 1}^K A_k\, \delta\left(\frac{t-t_k}{\td} \right),
  \label{eq:shot_noise_sde}
\end{equation}
where $\delta$ is the Dirac delta function. The model described by \Eqref{eq:shot_noise_sde} can be seen as a train of $\delta$-pulses, arriving at times $t_k$ with amplitudes $A_k$. These pulses pass through a filter causing an exponential decay with characteristic decay time $\td$, giving the process its name.

We remark that \Eqref{eq:shot_noise_sde} can be written as
\begin{equation}
  \Lin \Sn_K = f_K(t),
  \label{eq:shot_noise_sde_alt}
\end{equation}
where $\Lin$ is the linear operator $\Lin =1+ \td \, \D/\D t$ and the forcing is given by
\begin{equation}
  f_K(t) = \sum_{k=1}^K A_k \delta\left( \frac{t-t_k}{\td} \right). 
  \label{eq:f}
\end{equation}
A Green's function for the operator $\Lin$, that is any function fulfilling $\Lin G(t;s)=\delta(t-s)$, is given by
\begin{equation}
  G(t-s) = \exp\left( -\frac{t-s}{\td} \right) \Theta\left( \frac{t-s}{\td} \right).
  \label{eq:green}
\end{equation}
The solution of \Eqref{eq:shot_noise_sde_alt} is then the convolution
\begin{equation}
  \Sn_K(t) = \int_{-\infty}^\infty \D s\, G(t-s) f_K(s),
  \label{eq:shot_noise_green}
\end{equation}
where $G(t-s)$ can be interpreted as the filter which $f_K(t)$ passes through. Since both $G$ and $f_K$ are non-zero only for positive arguments, the integration in \Eqref{eq:shot_noise_green} could be taken over the interval $[0,t]$, and this is done in \Secref{sec:synth}. Here, however, we take the integration limits to infinity in order to remain consistent with the Fourier transforms, discussed below and in the appendices.

In \Appref{app:psd}, the power spectral density (PSD) of the FPP is shown to be
\begin{equation}
  \psd{\Sn} = \rSn^2 \frac{2 \td}{1+\td^2 \omega^2} + 2 \pi \m{\Sn}^2 \delta(\omega).
  \label{eq:psd_sn}
\end{equation}
The mean value of the signal gives a zero frequency contribution to the PSD, while the fluctuations around the mean value give rise to a Lorenzian power spectrum. Since the pulses are uncorrelated, there is no explicit dependence on the average waiting time $\tw$, apart from contributing to the value of $\tm{\Sn}$ and $\rSn$. Moreover, we see from \Eqsref{eq:psd_G} and \eqref{eq:psd_f} that the Poisson point process $f_K(t)$ provides the zero frequency contribution as well as a flat contribution independent from the frequency due to the lack of correlation between the pulses, while the Lorenzian spectrum comes entirely from the filter $G(t)$.

The auto-correlation and power spectral density are Fourier transform pairs under the Fourier transform over the entire real line. Thus we readily obtain the auto-correlation function
\begin{equation}
  R_\Sn(\tau) = \fourinv{S_\Sn} =  \frac{1}{2 \pi} \int\limits_{-\infty}^{\infty} \D \omega \, e^{ i \omega \tau } S_\Sn(\omega) = \rSn^2 \exp\left( -\frac{\abs{\tau}}{\td} \right) + \m{\Sn}^2,
  \label{eq:acorr_sn}
\end{equation}
where $\fourinv{\bullet}$ denotes the inverse Fourier transform. We see that the time dependence of the auto-correlation function comes entirely from the Green's function $G(t)$, as is expected, since the pulses are uncorrelated and thus the average time between pulses does not appear explicitly in the auto-correlation function.

\begin{figure}
  \centering
  \includegraphics[width = 0.6\textwidth]{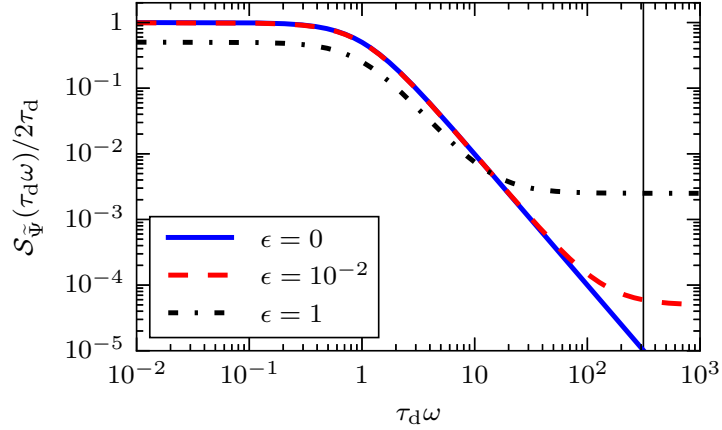}
  \caption{\label{fig:PSD_mult_eps}Comparison of power spectral densities with $\theta = 10^{-2}$ and various values of $\epsilon$. The solid line denotes both $\mathcal{S}_{\nSn}$ and $\mathcal{S}_{\norm{\dyn}}$ while the broken lines denote $\mathcal{S}_{\norm{\obs}}$. The vertical line gives the Nyquist frequency.}
\end{figure}

\begin{figure}
  \centering
  \includegraphics[width = 0.6\textwidth]{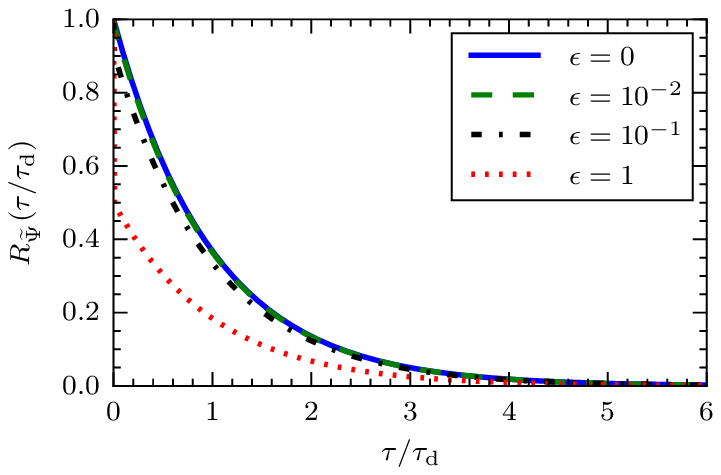}
  \caption{\label{fig:AC_mult_eps}Comparison of auto-correlation functions with $\theta = 10^{-2}$ and various values of $\epsilon$. The solid line denotes both $R_{\nSn}$ and $R_{\norm{\dyn}}$ while the broken lines denote $R_{\norm{\obs}}$.}
\end{figure}

It can be shown that the PSD and auto-correlation function of the normalized variable $\nSn$ are given by
\begin{align}
  \psd{\nSn} &= \frac{\psd{\Sn}-2 \pi \m{\Sn}^2 \delta(\omega)}{\rSn^2} =\frac{2 \td }{1+\td^2 \omega^2},\label{eq:psd-sn-norm}\\
  R_\nSn(\tau) &= \frac{R_\Sn(\tau)-\m{\Sn}^2}{\rSn^2} = \exp\left( -\frac{\abs{\tau}}{\td} \right). \label{eq:ac-sn-norm}
\end{align}
These expressions are presented in \Figsref{fig:PSD_mult_eps} and \ref{fig:AC_mult_eps}, respectively. The solid line in \Figref{fig:PSD_mult_eps} represents the PSD of $\nSn$, while the solid line in \Figref{fig:AC_mult_eps} represents the auto-correlation function of $\nSn$. The other elements in these figures will be discussed further in \Secref{sec:psd}. From \Eqref{eq:ac-sn-norm}, it is evident that the e-folding time of $R_\nSn$ is the pulse duration time $\td$. This corresponds to $\td \omega = 1$ in \Figref{fig:PSD_mult_eps}, giving the approximate frequency where the PSD changes from a flat spectrum to power law behavior.

\section{Additive noise}\label{sec:noise}
In this section, we will expand the FPP with two forms of additional noise terms, which will be referred to as either observational $(\obs)$ or dynamical $(\dyn)$ noise. In the following, the specific realization of the FPP (that is, the pulse amplitudes, arrival times and number of pulses for a particular realization) plays no role. Thus, for simplicity of notation, the subscript $K$ will be suppressed in the following.

Observational noise denotes noise unconnected to the FPP. Thus, a noise term is simply added to a realization of the process $\Sn(t)$;
\begin{equation}
  \obs(t) = \Sn(t) + \sigma N(t),
  \label{eq:shot_noise_obs}
\end{equation}
where $N(t)$ is a normally distributed process with vanishing mean and unit standard deviation and $\sigma$ is the noise intensity parameter, effectively describing the standard deviation of the noise process.

In the case of dynamical noise, the noise term is added as random forcing in the stochastic differential equation \eqref{eq:shot_noise_sde}, and is therefore connected to the pulses:
\begin{equation}
  \td \D \dyn = \left[ - \dyn(t) + \sum_{k = 1}^K A_k \delta\left( \frac{t-t_k}{\td} \right) \right] \D t  + \sqrt{2 \td} \sigma \D W
  \label{eq:shot_noise_dyn_sde}
\end{equation}
where $ W(t)$ is the Wiener process. It is possible to solve \Eqref{eq:shot_noise_dyn_sde} in the same way as \Eqref{eq:shot_noise_sde} was solved, giving
\begin{equation}
  \dyn(t) = \Sn(t) + \sigma Y (t),
  \label{eq:shot_noise_dyn}
\end{equation}
where 
\begin{equation}
  Y(t) = \sqrt{2/\td}  \int_{-\infty}^{\infty} G(t-s)\D W(s)
  \label{eq:shot_noise_dyn_green}
\end{equation}
is an Ornstein-Uhlenbeck process, and the random variable $Y$ is normally distributed with zero mean and unit standard deviation.

\begin{figure}
  \centering
    \includegraphics[width = 0.6\textwidth]{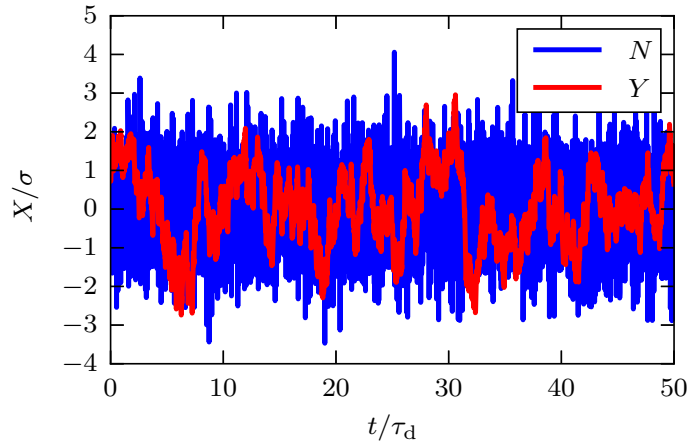}
    \caption{\label{fig:NvY}Synthetically generated observational and dynamic noise terms for $\theta = 10^{-2}$.}
\end{figure}

We also introduce a parameter $\epsilon$, describing the relative noise level. Using $X$ as a collective symbol for both $\sigma N$ and $\sigma Y$, we define $\epsilon$ as
\begin{equation}
  \epsilon = \left( \frac{X_\rms}{\rSn} \right)^2 = \frac{\sigma^2}{\gamma \mA^2}.
  \label{eq:eps_def}
\end{equation}
 Although $N(t)$ and $Y(t)$ have the same probability distributions, they exhibit very different dynamical behavior, as illustrated by realizations of the processes presented in \Figref{fig:NvY}. While $N(t)$ fluctuates rapidly on the sampling time scale around the zero value, $Y(t)$ wanders around the zero value, with finite temporal correlations. This is quantified in \Appref{app:psd_noise}, where the auto-correlation functions and power spectral densities of $N$ and $Y$ are derived. Note that both $N$ and $Y$ are independent of the process $\Sn$.

\begin{figure}
  \centering
  \includegraphics[width = 0.6\textwidth]{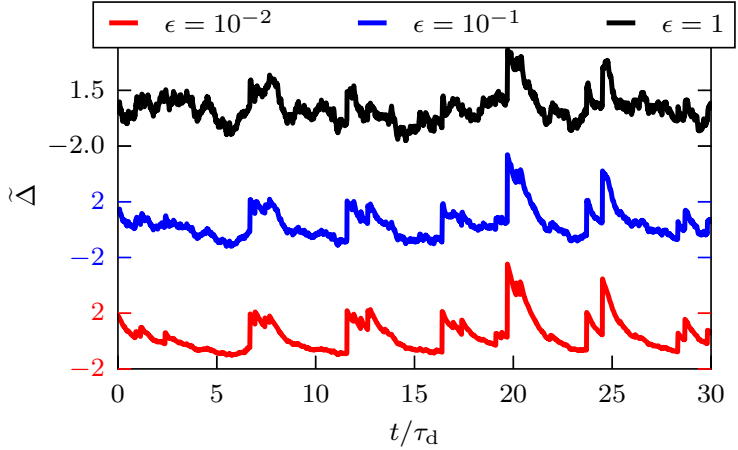}
  \caption{\label{fig:Sd_g1_dt0.01_mult_eps}Synthetically generated filtered Poisson processes with dynamical noise, $\gamma = 1$, $\theta = 10^{-2}$ and various values of $\epsilon$.}
\end{figure}

\begin{figure}
  \centering
  \includegraphics[width = 0.6\textwidth]{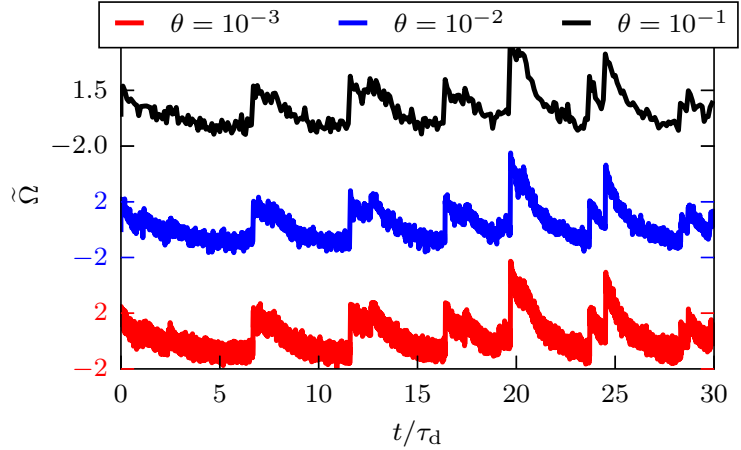}
  \caption{\label{fig:S_g1_e0.1_mult_dt}Synthetically generated filtered Poisson processes with observational noise, $\gamma = 1$, $\epsilon = 10^{-1}$ and various values of $\theta$.}
\end{figure}

Realizations of the processes $\Sn$, $\dyn$ and $\obs$ for different parameters $\gamma$, $\epsilon$ and $\theta$ are presented in \Figsref{fig:S_e0_dt0.01_mult_gamma}, \ref{fig:Sd_g1_dt0.01_mult_eps} and \ref{fig:S_g1_e0.1_mult_dt}. Realizations of the pure FPP for various $\gamma$ are presented in \Figref{fig:S_e0_dt0.01_mult_gamma}. For $\gamma < 1$, the signal is very intermittent and spends large amounts of time close to zero. For $\gamma > 1$, pulse overlap is much more significant, washing out the intermittent features. As $\gamma$ becomes very large, the signal resembles symmetric fluctuations around the mean value, and it can be shown that in the limit $\gamma \to \infty$, the filtered Poisson process has a probability distribution resembling a normal distribution \cite{garcia-prl-12,garcia-pop-16,rice-44,pecseli-fps}.

Realizations of the FPP with dynamical noise for various $\epsilon$ are presented in \Figref{fig:Sd_g1_dt0.01_mult_eps}. For very small $\epsilon$, there is very little difference between the pure process and the process with dynamical noise. For larger $\epsilon$, the noise process plays a larger role, concealing all but the largest pulses.

The effect of changing $\theta$ is clearest for the FPP with observational noise. Realizations of this process for various values of $\theta$ are presented in \Figref{fig:S_g1_e0.1_mult_dt}. Here we see that for large $\theta$, the process resembles a pure FPP. For small $\theta$ the noise process dominates, even though its rms-value is $1/10$th the rms-value of the FPP in this case. A smaller time step means more data points in a given time interval and thus more chances for large values of the noise process. In contrast, the pure FPP itself is much less sensitive to changes in time resolution, the primary effect being that separate pulses may be counted as one in the computation, if they are close enough. The FPP with dynamical noise is also less sensitive to $\theta$ due to the exponential damping the noise.

\section{Moments and distributions}\label{sec:momdist}

In this section, we present the lowest order moments and the PDF of the FPP with additive noise and describe how the model parameters can be estimated from these for a realization of the process.
The probability density function of the stationary process $\Sn$ is given in \Eqref{eq:sn_pdf}, and both $N$ and $Y$ are normally distributed with vanishing mean and unit standard deviation. Thus, we write
\begin{equation}
  \Snn = \Sn + X,
  \label{eq:process}
\end{equation}
where the random variable $X$ has the probability density function
\begin{equation}
  P_X(x; \sigma) = \frac{1}{\sqrt{2 \pi}\sigma} \exp\left( - \frac{x^2}{2 \sigma^2} \right).
  \label{eq:noise}
\end{equation}
Note that $X$ has the same distribution as both $\sigma N$ and $\sigma Y$, while $\Snn$ has the same distribution as both $\obs$ and $\dyn$. In other words, the PDF of the FPP with additional noise is the same for both types of noise considered here. In this section, we assume continuous random variables $X$ and $\Sn$. It can be shown that the convergence of the moments of $\Sn$ to their true values depends on $\theta$. A discussion on this and the estimation of the moments of $\Sn$ in general is given in \Refs{garcia-pop-16,kube-15}. Under the assumption that $\Sn$ and $X$ are independent, and using that $\tm{X}=0$, the lowest order moments of $\Snn$ are readily calculated as
\begin{subequations}
\begin{align}
  \m{\Snn} &= \m{\Sn+X} = \gamma \mA, \label{eq:mean-snn-start} \\
  \rSnn^2 &= \m{\left[ \left( \Sn + X \right)-\m{\Sn} \right]^2  } = \gamma \mA^2 + \sigma^2,\label{eq:rms-snn-start} \\
  S_{\Snn} &= \frac{\m{\left[ \left( \Sn + X \right)-\m{\Sn} \right]^3}}{\rSnn^3} =  \frac{2 \gamma \mA^3}{\left( \gamma \mA^2 + \sigma^2 \right)^{3/2}},\label{eq:skew-snn-start} \\
  F_{\Snn} &= \frac{\m{\left[ \left( \Sn + X \right)-\m{\Sn} \right]^4}}{\rSnn^4} = \frac{3}{\left( \gamma \mA^2 + \sigma^2 \right)^2} \left[ \gamma \mA^4 \left( \gamma + 2 \right) + 2 \gamma \mA^2 \sigma^2 + \sigma^4 \right].\label{eq:flat-snn-start}  
\end{align}
\end{subequations}
Using \Eqref{eq:eps_def}, we have the moments
\begin{subequations}
\begin{align}
  \m{\Snn} &= \gamma \mA,\label{eq:mean-snn} \\
  \Snn_\text{rms}^2 &= (1+\epsilon) \gamma \mA^2, \label{eq:rms-snn}\\
  S_{\Snn} &= \frac{2}{\left( 1+\epsilon \right)^{3/2} \gamma^{1/2}}, \label{eq:skew-snn}\\
  F_{\Snn} &= 3+ \frac{6}{\left(1+\epsilon\right)^2 \gamma}.\label{eq:flat-snn}
\end{align}
\end{subequations}
Also in this case, we can find a parabolic relation between the  skewness and flatness moments,
\begin{equation}
  F_{\Snn} = 3+\frac{3}{2} \left( 1+\epsilon \right) S_{\Snn}^2.
  \label{eq:parabolic-snn}
\end{equation}
The effect of additional noise is to increase the pre-factor in the parabolic relationship.

The model parameters $\gamma$ and $\epsilon$ can be estimated from the moments in a several different ways.  Using \Eqsref{eq:mean-snn} and \eqref{eq:rms-snn} to eliminate $\tm{A}$. The relative fluctuation level $\rSnn/\tm{\Snn}$ is then related to the model parameters by 
\begin{equation}
  \frac{\rSnn}{\m{\Snn}} = \sqrt{\frac{1+\epsilon}{\gamma}},
  \label{eq:rel_fluct}
\end{equation}
clearly showing how the additional noise amplifies the fluctuation level.
As the estimators for lower order moments are more accurate than those for higher order moments, it is reasonable to assume the most accurate estimators for the model parameters come from using the lowest order moments, $\rSnn/\tm{\Snn}$ and $S_{\Snn}$:
\begin{subequations}
\label{eq:par-est-rel-s}
\begin{align}
  \epsilon &= \left( \frac{2}{S_{\Snn}} \right)^{1/2} \left( \frac{\rSnn}{\m{\Snn}} \right)^{1/2} ,\label{eq:par-est-rel-s-e}\\
  \gamma &= \left( \frac{2}{S_{\Snn}} \right)^{1/2} \left( \frac{\m{\Snn}}{\rSnn} \right)^{3/2} .\label{eq:par-est-rel-s-g}
\end{align}
\end{subequations}
In experimental fluctuation data time series there can sometimes be reasons not to trust the mean value of a signal, for example due to externally imposed low frequency noise or trends \cite{garcia-pop-13,garcia-jnm-13}. Such problems typically do not affect the large-amplitude fluctuations, leaving the higher order moments trustworthy. In this case, using the normalization in \Eqref{eq:resc_sgnl} and observing that $S_{\nSnn} = S_{\Snn}$ and $F_{\nSnn} = F_{\Snn}$, we have 
\begin{subequations}
\label{eq:par-est-s-f}
\begin{align}
  \epsilon &= \frac{2}{3} \frac{ \left( F_{\nSnn}-3 \right) }{S_{\nSnn}^2} - 1,\label{eq:par-est-s-f-e}\\
  \gamma &= \frac{27}{2} \frac{S_{\nSnn}^4}{\left( F_{\nSnn}-3 \right)^3}.\label{eq:par-est-s-f-g}
\end{align}
\end{subequations}
In \Secref{sec:synth}, it will be shown that estimating the parameters from \Eqref{eq:par-est-rel-s} is preferable to using \Eqref{eq:par-est-s-f}, given that $\rSnn/\tm{\Snn}$ is reliable.\\

The probability density function of a sum of two independent random variables $\Sn$ and $X$ is a convolution of their respective probability density functions \cite{stark-prob}:
\begin{equation}
  P_\Snn(\snn;\gamma,\mA,\sigma) = \int_{-\infty}^{\infty}\text{d}\sn\, P_\Sn(\sn;\gamma,\mA) P_X(\snn-\sn;\sigma).
  \label{eq:PZstart}
\end{equation}
As a consistency check, it should be noted that in the limit $\epsilon \to 0$, $P_\Snn$ should be the probability density function of a pure FPP. This is indeed the case, as in this limit $\sigma \to 0$, and by definition,
\begin{equation*}
  \lim_{\sigma \to 0} P_X(\snn-\sn;\sigma) = \delta\left( \snn-\sn \right),
\end{equation*}
and the Gamma distribution of the FPP without noise is recovered. Thus, in the following, we take the case $\epsilon = 0$ to signify a FPP without additive noise. 

\begin{figure}
  \centering
    \includegraphics[width = 0.6\textwidth]{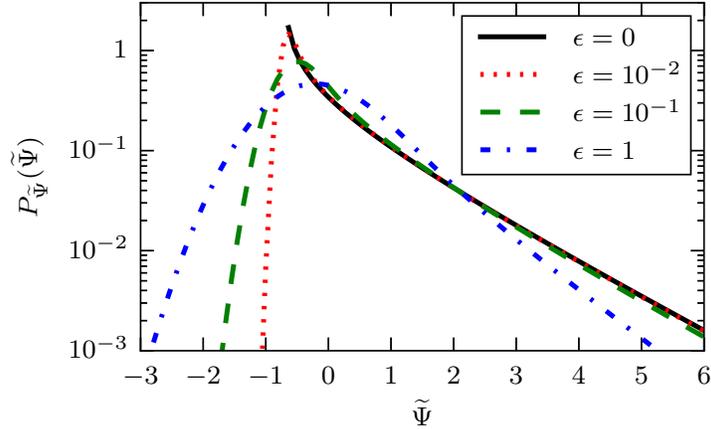}
    \caption{\label{fig:pdf_g0.5}Probability density function of the random variable $\Snn$ for $\gamma=1/2$ and various values of $\epsilon$.}
\end{figure}

\begin{figure}
  \centering
    \includegraphics[width = 0.6\textwidth]{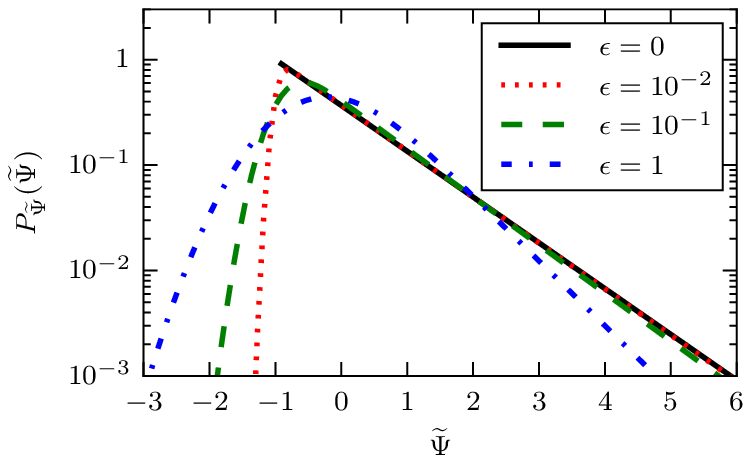}
    \caption{\label{fig:pdf_g1}Probability density function of the random variable $\Snn$ for $\gamma=1$ and various values of $\epsilon$.}
\end{figure}

\begin{figure}
  \centering
    \includegraphics[width = 0.6\textwidth]{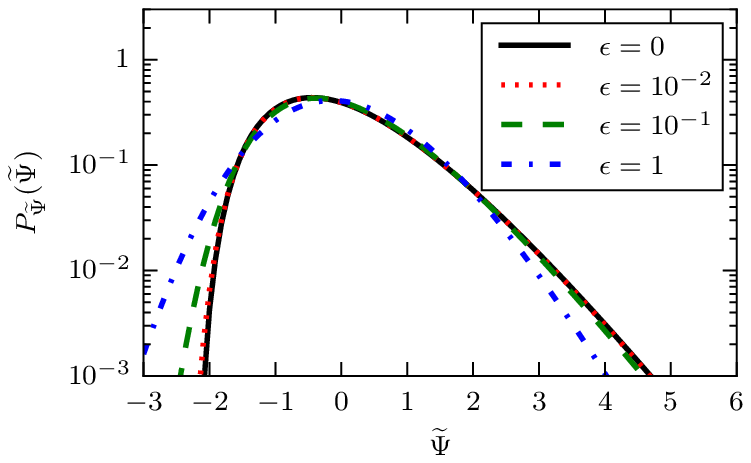}
    \caption{\label{fig:pdf_g5}Probability density function of the random variable $\Snn$ for $\gamma=5$ and various values of $\epsilon$.}
\end{figure}

The expression for the probability density function of $\nSnn$ is given in the appendix, \Eqref{eq:pdf_nsnn}. In order to illustrate the effect of pulse overlap and additional noise, the PDF for $\nSnn$ is shown for various values of $\epsilon$ and $\gamma = 1/2$, $1$ and $5$ in \Figsref{fig:pdf_g0.5}, \ref{fig:pdf_g1} and \ref{fig:pdf_g5}, respectively. Clearly, as $\epsilon$ increases beyond unity, the probability distribution changes towards a normal distribution. For $\epsilon = 0$, the random variable $\Snn$ is non-negative, causing an abrupt halt in the distribution for $\gamma \leq 1$. This jump does not exist for $\epsilon>0$. Thus negative values for $\Snn$, or equivalently, $\nSnn < -\gamma^{1/2}$, signifies a deviation from a pure FPP.


\section{Spectra and correlations}\label{sec:psd}
The PSD of the sum of two independent random variables is the sum of their respective PSDs. The power spectra for the FPP with additional noise are derived in \Appref{app:psd}. With \Eqsref{eq:psd_sn} and \eqref{eq:psd_Y}, the power spectral density of the FPP with dynamical noise is therefore
\begin{equation}
  \psd{\dyn} = \psd{\Sn} + \sigma^2 \psd{Y} = \dyn_\rms^2 \frac{2\td}{1+ \td^2 \omega^2} + 2 \pi \m{\dyn}^2 \delta(\omega),
  \label{eq:psd_dyn}
\end{equation}
where $\tm{\dyn}$ and $\dyn_\rms$ are given by \Eqsref{eq:mean-snn} and \eqref{eq:rms-snn}, respectively. We have also used the relation $\sigma^2 = \epsilon \gamma \mA^2$. Note that the spectrum in \Eqref{eq:psd_dyn} is identical to that for the pure FPP. Thus, we conclude that the auto-correlation functions of $\Sn$ and $\dyn$ have the same functional shape as well, up to the difference in the first two moments.

The power spectral density of the FPP with observational noise is obtained from \Eqsref{eq:psd_sn} and \eqref{eq:psd_N_discrete}:
\begin{equation}
  \psd{\obs} =  \obs_\rms^2 \frac{2 \td}{1+\epsilon} \left[ \frac{1}{1+\td^2 \omega^2} + \frac{\epsilon}{2} \theta  \right] + 2 \pi \m{\obs}^2 \delta(\omega).
  \label{eq:psd_obs}
\end{equation}
This function is qualitatively different from the power spectral densities of $\Sn$ and $\dyn$, although it converges to both in the limit $\epsilon \to 0$. These differences are now explored in detail.

\begin{figure}
  \centering
  \includegraphics[width = 0.6\textwidth]{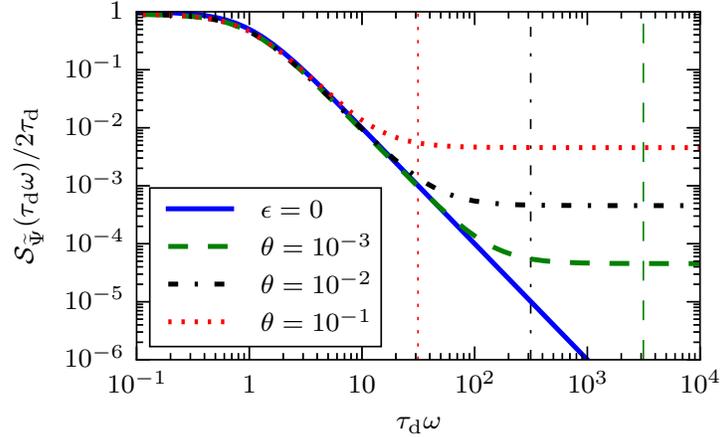}
  \caption{\label{fig:PSD_mult_theta}Comparison of power spectral densities with $\epsilon = 10^{-2}$ and various values of $\theta$. The solid line denotes $\mathcal{S}_{\nSn}$ and $\mathcal{S}_{\norm{\dyn}}$ while the broken lines denote $\mathcal{S}_{\norm{\obs}}$. The vertical lines give the Nyquist frequency of the corresponding expressions.}
\end{figure}

Using the normalizations in \Eqsref{eq:resc_sgnl} and \eqref{eq:psd-sn-norm}, we can list the power spectral densities of the rescaled signals $\nSn$, $\norm{\obs}$ and $\norm{\dyn}$ as
\begin{subequations}
\begin{align}
  \psd{\nSn} &= \psd{\norm{\dyn}} = 2 \td \frac{1}{1+\td^2 \omega^2} \label{eq:psd_sn_dyn_resc} \\
  \psd{\norm{\obs}} &= \frac{2 \td}{1+\epsilon} \left[ \frac{1}{1+\td^2 \omega^2} + \frac{\epsilon}{2}\theta \right]. \label{eq:psd_obs_resc}
\end{align}
\end{subequations}
The power spectral density of $\nSnn$ is presented in \Figsref{fig:PSD_mult_eps} and \ref{fig:PSD_mult_theta}. In \Figref{fig:PSD_mult_eps}, the difference between $\nSn$ or $\norm{\dyn}$ and $\norm{\obs}$ is presented. Higher ratio of noise signal to FPP decreases the value of the power spectral density for low frequencies and causes a transition from a power law spectrum to a constant spectrum at higher frequencies. The Nyquist frequency $\omega_\text{N}$ for $\theta = 10^{-2}$ is indicated by the vertical line,
\begin{equation}
  \omega_\text{N} = 2 \pi \frac{1}{2 \dt} = \frac{\pi}{\td \theta},
  \label{eq:nyquist_freq}
\end{equation}
and shows that for small $\epsilon$, this transition happens at too high frequencies to be reliablely observed. In \Figref{fig:PSD_mult_theta}, the power spectral density of $\norm{\obs}$ is presented for $\epsilon = 10^{-1}$ and various values of $\theta$. For low frequencies, the difference between $\psd{\nSn}$ and $\psd{\norm{\obs}}$ is too small to be of practical use, while the effect of noise for high frequencies can only be observed without aliasing for very low $\theta$.

The auto-correlation function of $\Sn$ is given in \Eqref{eq:acorr_sn}. Since the power spectral density of the rescaled processes $\nSn$ and $\norm{\dyn}$ are identical, we conclude that their auto-correlation functions are as well, 
\begin{equation}
  R_{\nSn}(\tau) = R_{\norm{\dyn}}(\tau) = \exp\left( -\frac{\abs{\tau}}{\td} \right),
  \label{eq:norm_acorr_Sn_dyn}
\end{equation}
where the auto-correlation of a rescaled random process is 
\begin{equation*}
  R_{\nSnn}(\tau) = \frac{R_{\Snn}(\tau) - \m{\Snn}^2}{\rSnn^2}.
\end{equation*}
With the auto-correlation function of $N(t)$ from \Eqref{eq:acorr_N}, we have that the auto-correlation function of $\norm{\obs}$ is
\begin{equation}
 R_\norm{\obs}(\tau)= \frac{1}{1+\epsilon} \left[ \exp\left( -\frac{\abs{\tau}}{\td} \right) + \epsilon \left( 1-\frac{\abs{\tau}}{\theta \td} \right) \Theta\left( 1-\frac{\abs{\tau}}{\theta \td} \right)  \right].
  \label{eq:acorr_obs_norm}
\end{equation}
Note that $R_{\norm{\obs}}(0)=1$ as required by the normalization, while for correlation times longer than the sampling time, the correlation function of the normalized variable $\norm{\obs}$ has a value of $1/(1+\epsilon)$ times the correlation function of $\norm{\dyn}$. Thus, \Eqref{eq:acorr_obs_norm} can also be written as
\begin{equation}
  R_\norm{\obs}(\tau) = \begin{cases}
    1, & \tau = 0 \\
    \frac{1}{1+\epsilon} \exp\left( - \frac{ \abs{\tau}}{\td} \right), & \abs{\tau} \geq \dt \end{cases}
  \label{eq:acorr_obs_norm_2}
\end{equation}
If there is no appreciable drop from $\tau = 0$ to $\tau = \dt$, and the correlation functions overlap for $\tau \geq \dt$, the observational noise is negligible.

\begin{figure}
  \centering
  \includegraphics[width = 0.6\textwidth]{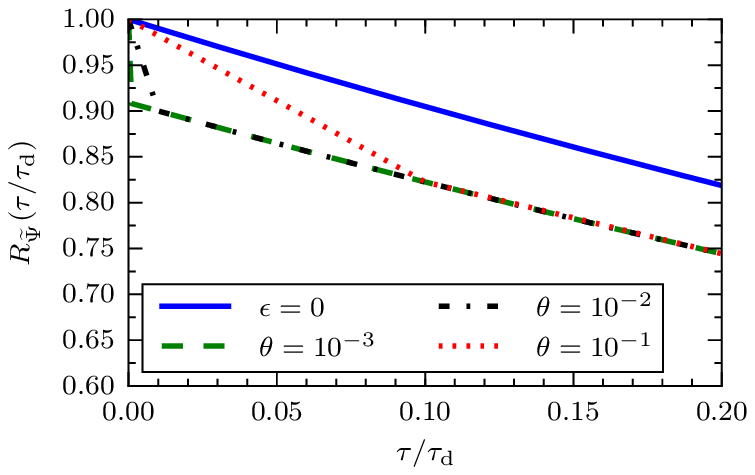}
  \caption{\label{fig:AC_mult_dt}Comparison of auto-correlation functions with $\epsilon = 10^{-1}$ and various values of $\theta$. The solid line denotes both $R_{\nSn}$ and $R_{\norm{\dyn}}$, while the broken lines denote $R_{\norm{\obs}}$.}
\end{figure}

As expected, in the limit $\epsilon \to 0$, $R_\norm{\obs}$ approaches $R_\nSn$. In practice, the auto-correlation function should be better at revealing the presence of observational noise than the power spectral density, as this difference is largest for small time lags, where the auto-correlation function is the most accurate. Still, $R_\norm{\dyn}(\dt) = \exp(-\theta)$ and $R_\norm{\obs}(\dt) = \exp(-\theta)/(1+\epsilon)$, a difference which may be difficult to verify in practice for small, but appreciable $\epsilon$ (say, $\epsilon \sim 10^{-1}$). An example of this is seen in \Figref{fig:AC_mult_eps}. In this figure, the auto-correlation function of $\nSnn$ is presented as a function of $\tau/\td$ for $\theta = 10^{-2}$ and various values of $\epsilon$. For large $\epsilon$, both the initial drop and the reduced value of $R_{\norm{\obs}}$ compared to $R_{\norm{\dyn}}$ is clear, but this is not the case for $\epsilon\leq 10^{-1}$. The behavior of the auto-correlation of $\norm{\obs}$ for $\epsilon = 10^{-1}$ and varying $\theta$ is shown in \Figref{fig:AC_mult_dt}. All functions fulfill $R_{\norm{\Snn}}(0) = 1$, and for $\tau>\dt$, all $R_\norm{\obs}$ have the same value.

\section{Parameter estimation}\label{sec:synth}
In this section, we present results from numerical analysis using synthetically generated time series. The time series are generated following the convolutions in \Eqsref{eq:shot_noise_green} and \eqref{eq:shot_noise_dyn_green}, with integration limits $[0,t]$. The convolutions are performed by a fast Fourier transform numerical convolution. All random numbers are generated using a Mersenne Twister. The time array is constructed as $t_m =m \dt$, with $m = 0,1,\cdots,M-1$ and $T = \dt M$. We have set $\td = 1$, so we vary $\theta$ by varying $\dt$. The $K$ pulse amplitudes are drawn from an exponential distribution with $\mA=1$, the $K$ arrival times correspond to $K$ integers $\left\{ m_k \right\}_{k=1}^{K}$ uniformly distributed on $[0,M-1]$, giving
\begin{equation}
  f_K(t) \rightarrow f_K[m] = \sum_{k=1}^{K} A_k \delta_{m m_k}
  \label{eq:f_gen}
\end{equation}
where $\delta_{a b}$ is the Kronecker delta function for integers $a$ and $b$. $N[m]$ is an array of $M$ independent and identically distributed normal variables with vanishing mean and unit standard deviation, while the discrete version of $\D W$ is an equally shaped array of independent and identically distributed normal variables with vanishing mean and standard deviation $\dt^{1/2}$.

\subsection{Reliability of parameter estimation}
The aim of this section is to numerically test and verify results from the prior sections. We present analysis of parameter estimation for synthetic time series with the goal of separating the two types of noise, where the analysis is performed on 1000 time series of each type, each of length $M=10^6$ data points, with parameters close to experimental values (see Refs.~\onlinecite{kube-16,theodorsen-ppcf-16,garcia-nf-07}); $\gamma = 2$, $\epsilon = 5\times10^{-2}$ and $\theta = 2 \times 10^{-2}$. This gives time series of duration $T/\td = 2 \times 10^{4}$.

Given a time series, we can estimate $\gamma$ and $\epsilon$ by comparing the estimated PDF for the time series to the PDF in \Eqref{eq:pdf_nsnn}, or by comparing the sample estimate of the moments to the moments in \Eqsref{eq:par-est-rel-s} or \eqref{eq:par-est-s-f}. These two methods do not discriminate the nature of the noise. However, comparing \Eqsref{eq:norm_acorr_Sn_dyn} and \eqref{eq:acorr_obs_norm_2}, we see that \Eqref{eq:acorr_obs_norm_2} gives the auto-correlation function of a filtered Poisson process with observational noise for $\epsilon>0$ and the auto-correlation function of a filtered Poisson process with dynamical noise or without noise for $\epsilon = 0$. Thus \Eqref{eq:acorr_obs_norm_2} can potentially separate observational from dynamical noise. In addition, it gives us an estimate of $\theta$. The same considerations hold for the power spectral densities in \Eqref{eq:psd_sn_dyn_resc} as compared to \Eqref{eq:psd_obs_resc}. In all the following cases where we fit synthetic data to an analytical function, the initial values are given as the true parameter values of the process.

\begin{figure}
  \centering
  \includegraphics[width = 0.6\textwidth]{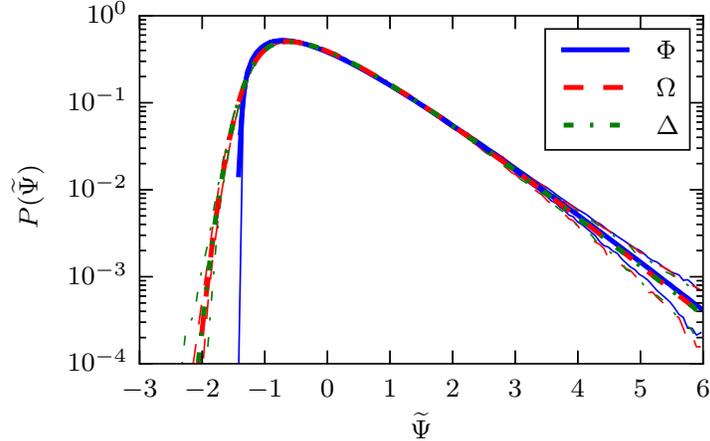}
  \caption{\label{fig:pdf_synth} Mean probability density functions (thick lines) and the maximal deviations from the mean (thin lines) of synthetically generated signals for $\gamma = 2$, $\epsilon = 5 \times 10^{-2}$, $\theta = 2 \times 10^{-2}$, and $10^3$ samples, each with $10^6$ data points.}
\end{figure}

\begin{figure}
  \centering
  \begin{subfigure}{0.47\textwidth}
    \includegraphics[width = \textwidth]{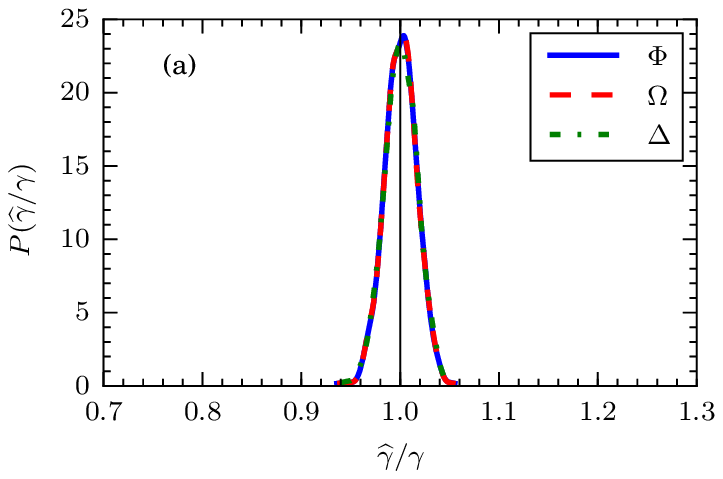}
    \caption{\label{fig:moment_mrs_synth_g}PDFs of $\widehat{\gamma}/\gamma$ from the relative fluctuation level and skewness.}
  \end{subfigure}
  ~
  \begin{subfigure}{0.47\textwidth}
    \includegraphics[width = \textwidth]{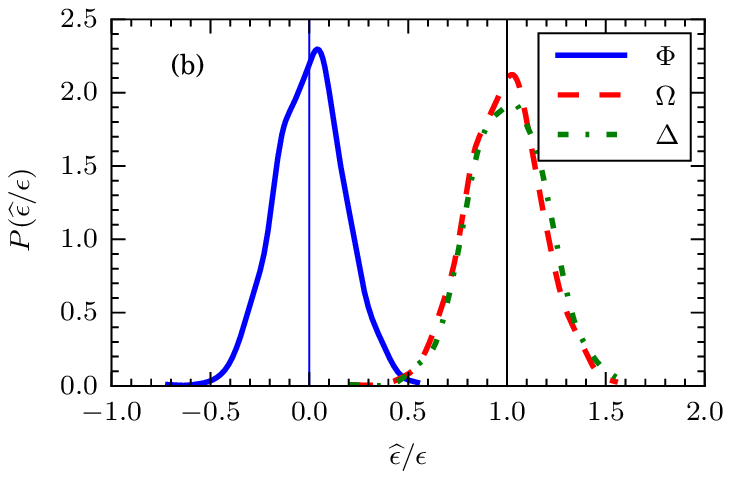}
    \caption{\label{fig:moment_mrs_synth_e}PDFs of $\widehat{\epsilon}/\epsilon$ from the relative fluctuation level and skewness.}
  \end{subfigure}

  \begin{subfigure}{0.47\textwidth}
    \includegraphics[width = \textwidth]{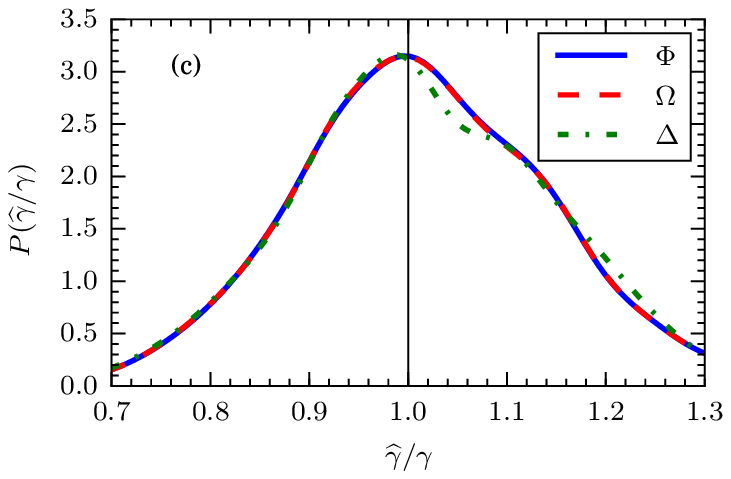}
    \caption{\label{fig:moment_sf_synth_g}PDFs of $\widehat{\gamma}/\gamma$ from the relative skewness and flatness moments.}
  \end{subfigure}
  ~
  \begin{subfigure}{0.47\textwidth}
    \includegraphics[width = \textwidth]{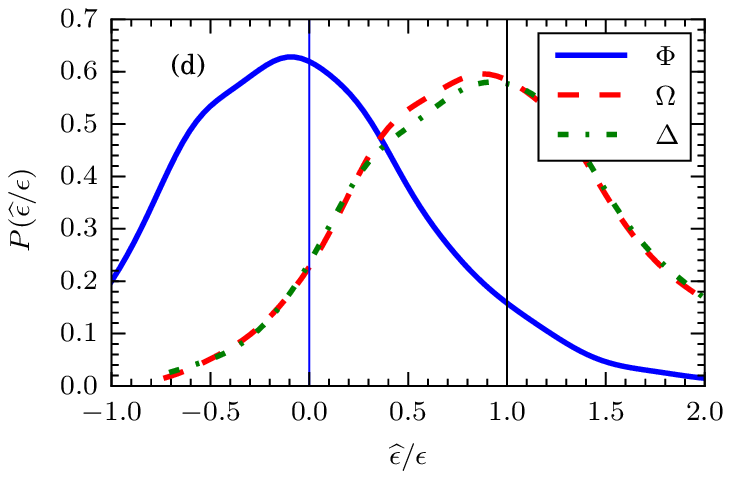}
    \caption{\label{fig:moment_sf_synth_e}PDFs of $\widehat{\epsilon}/\epsilon$ from the relative skewness and flatness moments.}
  \end{subfigure}

  \begin{subfigure}{0.47\textwidth}
    \includegraphics[width = \textwidth]{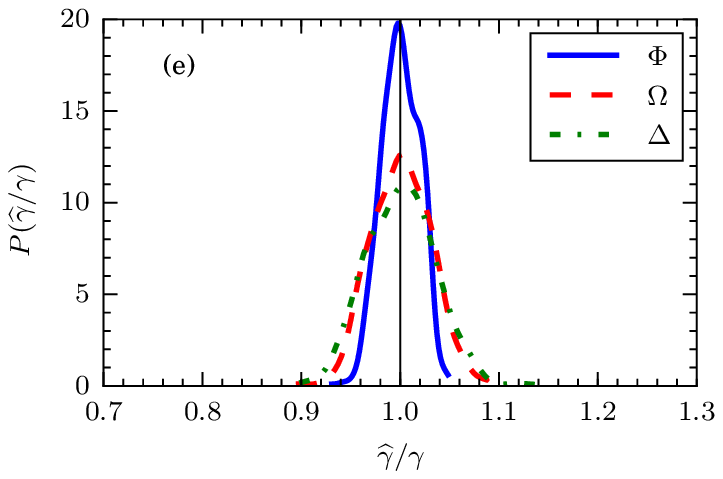}
    \caption{\label{fig:pdf_synth_g}PDFs of $\widehat{\gamma}/\gamma$ from fitting to the PDF.}
  \end{subfigure}
  ~
  \begin{subfigure}{0.47\textwidth}
    \includegraphics[width = \textwidth]{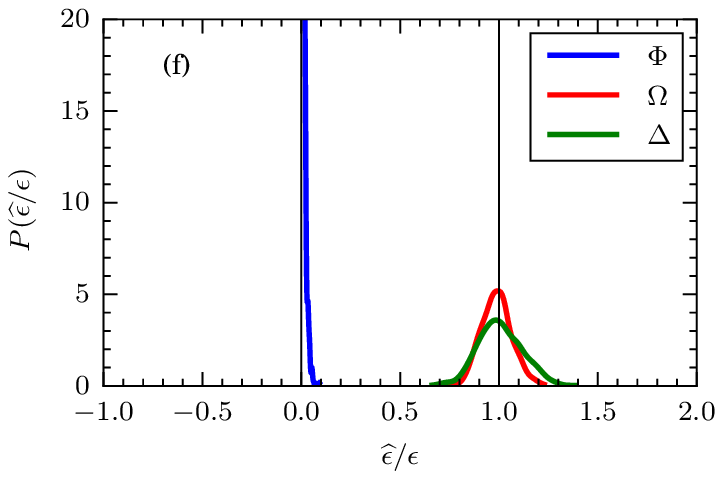}
    \caption{\label{fig:pdf_synth_e}PDFs of $\widehat{\epsilon}/\epsilon$ from fitting to the PDF.}
  \end{subfigure}
  \caption{\label{fig:synth_g_e_est} PDFs of the intermittency parameter $\widehat{\gamma}/\gamma$ and the noise level parameter $\epsilon$ estimated from the moments and the PDF of synthetically generated signals. The thin vertical lines give the true parameters.}
\end{figure}

The estimated PDFs of the synthetic signals are presented in \Figref{fig:pdf_synth}. The thick lines give the average PDF for all synthetic signals, and the thin lines indicate the maximal deviation from this mean value. Visually, the only reliable difference between the signals is the elevated tail for negative values of $\norm{\dyn}$ and $\norm{\obs}$ compared to $\nSn$. In \Figref{fig:synth_g_e_est}, we present PDFs for the estimated parameters $\widehat{\gamma}$ and $\widehat{\epsilon}$ from the moments and the PDF of the processes. Here and in the following, the hat symbol $\widehat{\bullet}$ indicates an estimated value. The values in \Figsref{fig:moment_mrs_synth_g} and \ref{fig:moment_mrs_synth_e} were obtained by estimating the relative fluctuation level and skewness of the synthetic data and using \Eqref{eq:par-est-rel-s}. The values in \Figsref{fig:moment_sf_synth_g} and \ref{fig:moment_sf_synth_e} were obtained by estimating the skewness and flatness and using \Eqref{eq:par-est-s-f}. In \Figsref{fig:pdf_synth_g} and \ref{fig:pdf_synth_e}, the values were obtained by fitting the function in \Eqref{eq:pdf_nsnn} to the estimated PDF of the synthetic data with a least squares routine under the constraints $\widehat{\gamma} \geq 10^{-1}$ and $\widehat{\epsilon} \geq 10^{-6}$ to ensure convergence. The rms-value of the distributions of $\widehat{\gamma}$ and $\widehat{\epsilon}$ are presented in Tables \ref{table:g-rms} and \ref{table:e-rms}, respectively.

As is evident from \Figref{fig:synth_g_e_est}, using the three lowest order moments or the PDF of the signal to estimate the parameters is far preferable to using the estimated skewness and flatness moments of the signal. \Figsref{fig:moment_sf_synth_g} and \ref{fig:moment_sf_synth_e} have far broader distributions than the other methods of parameter estimation. Due to the significant overlap between the distributions in \Figref{fig:moment_sf_synth_e}, the presence or absence of noise can be difficult to determine reliably. For the underlying parameters used here, Table \ref{table:g-rms} shows that the three lowest order moments give a better estimate for $\gamma$ than the PDF, while Table \ref{table:e-rms} shows the reverse for $\epsilon$. This result is likely highly dependent on the estimation methods employed, as well as the properties of the time series in question. Thus, the authors recommend a full Monte Carlo analysis, as presented here, in order to determine errors in parameter estimation.  As expected, the type of noise cannot be determined from the moments or PDF of the signal.

\begin{table}
\begin{tabular}{c | c | c | c}
  $\left( \widehat{\gamma}/\gamma \right)_\rms$ & $\rSnn/\m{\Snn}$, $S_\Snn$ & $S_\Snn$, $F_\Snn$ & $P(\nSnn)$ \\
  \hline
  $\Sn$ & 0.016 & 0.13 & 0.019 \\
  $\dyn$ & 0.017 & 0.13 & 0.030 \\
  $\obs$ & 0.016 & 0.13 & 0.035
\end{tabular}
\caption{\label{table:g-rms} Standard deviation of estimated $\gamma$-values of synthetically generated signals for $\gamma = 2$, $\epsilon = 5 \times 10^{-2}$, $\theta = 2 \times 10^{-2}$, and $10^3$ samples, each with $10^6$ data points.}
\end{table}

\begin{table}
\begin{tabular}{c | c | c | c}
  $\left( \widehat{\epsilon}/\epsilon \right)_\rms$ & $\rSnn/\m{\Snn}$, $S_\Snn$ & $S_\Snn$, $F_\Snn$ & $P(\nSnn)$ \\
  \hline 
  $\Sn$ & 0.17 & 0.64 & 0.011 \\
  $\dyn$ & 0.19 & 0.69 & 0.077 \\
  $\obs$ & 0.19 & 0.67 & 0.11
\end{tabular}
\caption{\label{table:e-rms} Standard deviation of estimated $\epsilon$-values of synthetically generated signals for $\gamma = 2$, $\epsilon = 5 \times 10^{-2}$, $\theta = 2 \times 10^{-2}$, and $10^3$ samples, each with $10^6$ data points.}
\end{table}

\begin{figure}
  \centering
  \begin{subfigure}{0.47\textwidth}
    \includegraphics[width = \textwidth]{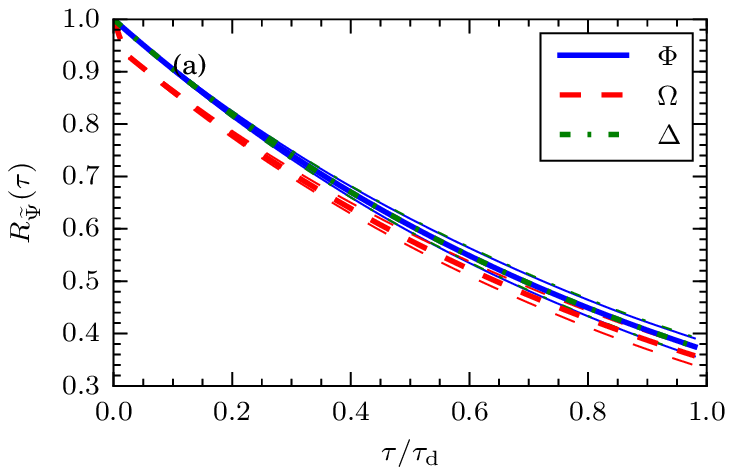}
    \caption{\label{fig:ac_synth}Mean auto-correlation functions (thick lines) and the maximal deviations from the mean (thin lines).}
  \end{subfigure}
  ~
  \begin{subfigure}{0.47\textwidth}
    \includegraphics[width = \textwidth]{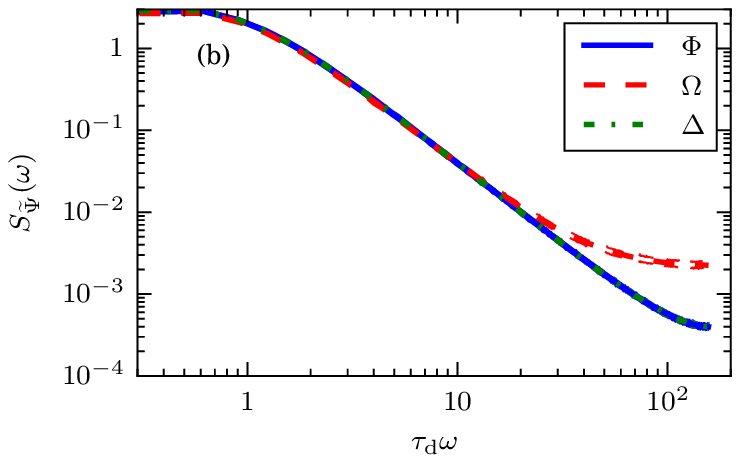}
    \caption{\label{fig:psd_synth}Mean power spectral density (thick lines) and the maximal deviations from the mean (thin lines).}
  \end{subfigure}

  \begin{subfigure}{0.47\textwidth}
    \includegraphics[width = \textwidth]{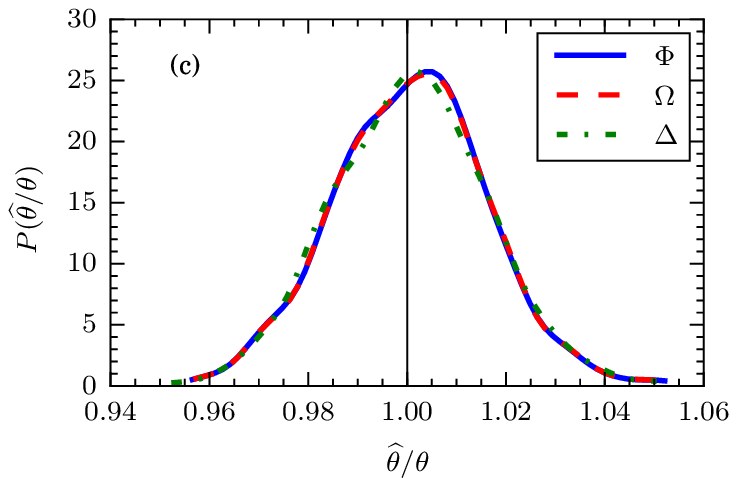}
    \caption{\label{fig:ac_synth_td}PDFs of $\widehat{\theta}/\theta$ from fitting to the auto-correlation function.}
  \end{subfigure}
  ~
  \begin{subfigure}{0.47\textwidth}
    \includegraphics[width = \textwidth]{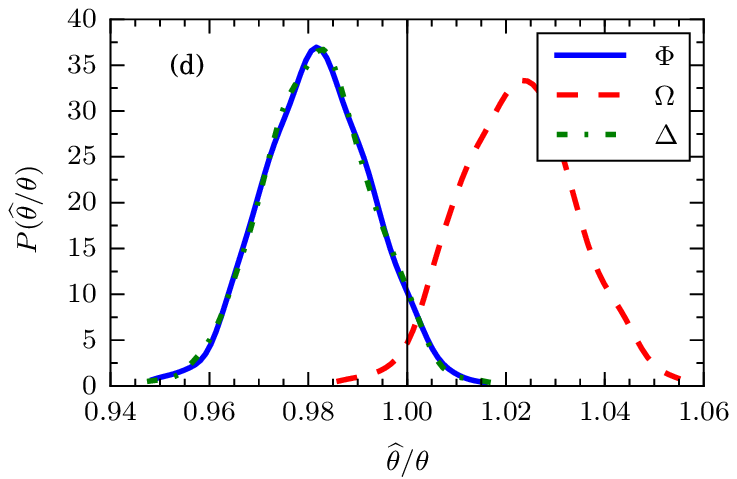}
    \caption{\label{fig:psd_synth_td}PDFs of $\widehat{\theta}/\theta$ from fitting to the power spectral density.}
  \end{subfigure}

  \begin{subfigure}{0.47\textwidth}
    \includegraphics[width = \textwidth]{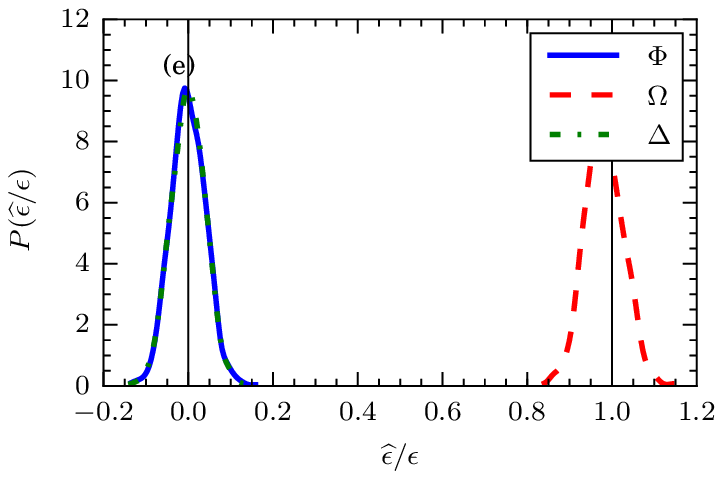}
    \caption{\label{fig:ac_synth_e}PDFs of $\widehat{\epsilon}/\epsilon$ from fitting to the auto-correlation function.}
  \end{subfigure}
  ~
  \begin{subfigure}{0.47\textwidth}
    \includegraphics[width = \textwidth]{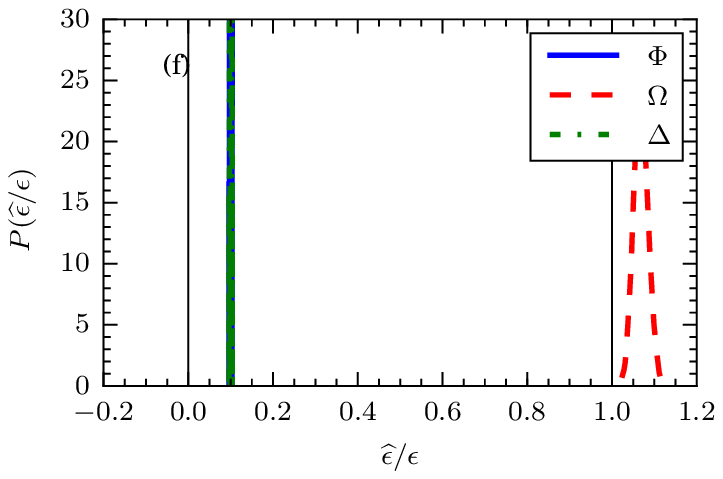}
    \caption{\label{fig:psd_synth_e}PDFs of $\widehat{\epsilon}/\epsilon$ from fitting to the power spectral density.}
  \end{subfigure}
  \caption{\label{fig:ac_psd_synth_theta_e_est} Results from the auto-correlation function and power spectral density of synthetically generated signals with $\gamma = 2$, $\epsilon = 5 \times 10^{-2}$, $\theta = 2 \times 10^{-2}$ and samples with $10^6$ data points. The thin vertical lines give the true parameters.}
\end{figure}

The auto-correlation functions of the synthetic signals are presented in \Figref{fig:ac_synth}, while the PSDs are presented in \Figref{fig:psd_synth}. Again the thick lines give the average result from all signals, while the thin lines give the maximal deviation from the average. The PSD was computed using Welch's method with $2^{10}$ data points window size, $50\%$ overlap and a Hanning window. In both cases, the observational noise gives a clear visual difference compared to the pure FPP and the same process with dynamical noise, having an elevated tail as predicted in \Eqref{eq:psd_obs_resc}. Note also the slight lifting of the tail in \Figref{fig:psd_synth} for $\Sn$ and $\dyn$. This is most likely caused by roundoff errors near the Nyquist frequency. Changing the window size or the windowing function does not correct the problem. Direct computation of the periodogram using a fast Fourier transform does not have this problem, although it presents other problems for parameter estimation. The effect of this lifting of the tail on parameter estimation is discussed below.

In order to estimate $\theta$ and $\epsilon$, we have fitted the auto-correlation function of the synthetic signals to the discrete version of \Eqref{eq:acorr_obs_norm_2}:
\begin{equation}
  R[m] = \displaystyle{\frac{1}{1+\epsilon}} \exp(- \theta m),\quad  m \geq 1,
  \label{eq:acorr-obs-norm-discrete}
\end{equation}
where we ignore the $m=0$ contribution, since this equals unity for all processes discussed. Due to the uncertainties in the auto-correlation function for large time lags, only the first 50 time steps are used. The PSD of the synthetic signals is fitted to the function in \Eqref{eq:psd_obs_resc}. In both cases, a non-linear least-squares fit routine with the true values as initial values was used.
In \Figsref{fig:ac_synth_td} and \ref{fig:ac_synth_e} the PDFs of $\widehat{\theta}$ and $\widehat{\epsilon}$ from the auto-correlation are presented, respectively. As $\theta$ is the same for all three classes of signal, the over-lapping PDFs are no surprise, and we note the small variation around the true value ($\widehat{\theta}_\rms = 0.015 \theta$ for all processes). Concerning $\widehat{\epsilon}$, the auto-correlation function does indeed separate the observational noise from the dynamical noise. For the observational noise, the rms-value is low, $\widehat{\epsilon}_\rms = 0.046 \epsilon$, and the auto-correlation function can be used to estimate the noise level. The PDFs of $\widehat{\theta}$ and $\widehat{\epsilon}$ from the PSD are presented in \Figsref{fig:psd_synth_td} and \ref{fig:psd_synth_e}, respectively. Again the rms-values of all parameters is small, $1-2\%$ of the original parameters, with the exception of the $\epsilon$--parameter in \Figref{fig:psd_synth_e} for $\Sn$ and $\dyn$, with $\widehat{\epsilon}_\rms = 0.0012 \epsilon$. In these figures, we have a clear bias in both $\widehat{\theta}$ and $\widehat{\epsilon}$. This is most likely created by the lifting of the tail of the power spectral density estimate, as discussed above. An elevated tail signifies observational noise, so an artificial elevated tail leads to an over-estimation of $\epsilon$, as seen in \Figref{fig:psd_synth_e}. This leads to a bias in the estimation of $\theta$, as seen in \Figref{fig:psd_synth_td}. Eliminating this bias by restricting the fit range is not recommended, as this will compromise the accuracy in separating the PSDs. As this bias is reproducible, it does not present a significant problem for parameter estimation, and the power spectral density could be used as a sanity check for the auto-correlation function.

\subsection{Level crossing statistics}
\begin{figure}
  \centering
  \includegraphics[width = 0.6\textwidth]{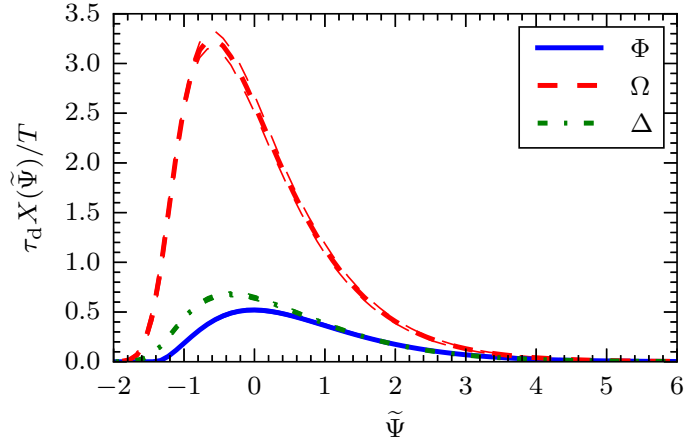}
  \caption{\label{fig:level_cross}Mean rate of level crossings (thick lines) and the maximal deviation from the mean (thin lines) for synthetically generated time series with $\gamma = 2$, $\epsilon = 5 \times 10^{-2}$ and $\theta = 2 \times 10^{-2}$. }
\end{figure}

Another measure which would intuitively separate the two types of noise is the number of upwards crossings above a certain threshold level per unit time, or the rate of level crossings. This quantity has been explored in \Refs{rice-45, theodorsen-pop-16,bierme-12,israel-72,garcia-pop-16} for the pure FPP, in \Refs{barakat-88,hopcraft-07} for a gamma distributed random process and in \Refs{sato-12,fattorini-12} for atmospheric plasma. The rate of level crossings above a threshold $\Snn$ as a function of the threshold is presented in \Figref{fig:level_cross} for the synthetic data discussed in the previous section. The thick lines give the mean values for the given threshold, while the thin lines represent the minimal and maximal value for all synthetic time series generated. The threshold is in units of signal rms-value above signal mean value. In agreement with intuition, the FPP with observational noise crosses the threshold much more frequently than the two others due to the rapid fluctuations around the mean value of the pure FPP at any amplitude triggering spurious crossings. The difference between the pure FPP and the FPP with dynamical noise is largest for small threshold values, where the number of threshold crossings is largest. Note that while $\Sn$ has its maximum number of level crossings for $\nSnn = 0$, this is down-shifted for the processes with noise, since the noise does not contribute to the mean value of the process. While we know of no theoretical estimate for the rate of level crossings for a FPP with noise, this value can still be found from synthetic signals, generated by estimating $\gamma$ and $\epsilon$ from the PDF of a measurement signal and $\theta$ from its auto-correlation function. Comparing the true rate of level crossings to the rate of level crossings for synthetic signals with different types of noise could separate the noise types. 

\section{Conclusions}\label{sec:conclusion}
Motivated by previous analysis of measurement data from magnetically confined plasmas, we have here investigated a FPP with normally distributed noise included as either observational or dynamical noise. The PDF, PSD and auto-correlation function of the pure FPP have been presented. These have also been extended to include noise, showing how the gamma probability distribution of the FPP moves towards a normal distribution as the influence of noise increases and how dynamical noise gives the same auto-correlation and PSD as the pure FPP. The model including noise has a PDF in accordance with recent measurements \cite{kube-16,garcia-kstar} and a parabolic relation between the skewness and flatness moments, as found in a variety of experiments on magnetized plasmas \cite{kube-16,theodorsen-ppcf-16,garcia-nf-07,sattin,garcia-pop-13,garcia-jnm-13}. The model also has an approximately exponentially decaying auto-correlation function and a power law PSD, in accordance with measurements from the scrape-off layer of a range of fusion plasmas \cite{theodorsen-ppcf-16,garcia-nf-07,kube-16,cziegler-10,dewhurst-08,tanaka-09}. The previous analysis of experimental data demonstrate that significant noise levels can be present, and the extension of the FPP presented here is indeed necessary in order to reliably estimate the underlying parameters of the process.

Using synthetically generated time series with experimentally relevant parameters, we have shown that the PDF of the FPP with additional noise is indeed capable of separating a process with noise from a noise-less process, and the sample variance of the estimated parameters provide an indication of the error of estimation. In addition, it was shown that estimating the intermittency parameter and the noise parameter from the estimated relative fluctuation level and skewness of the process is comparable in accuracy to estimating the parameters from the estimated PDF of the process, while using the estimated skewness and flatness gives a much worse estimate. It was furthermore shown that the auto-correlation function can provide an estimate of the characteristic decay time of the FPP, and is also capable of separating the types of noise. In the case of observational noise, the auto-correlation function can also provide an estimate of the noise level, but can not do this in the case of dynamical noise. The PSD has comparable capabilities to the auto-correlation function, but care must be taken in its use, in particular in the case of marginal sampling frequency. Lastly, the rate of level crossings was proposed as another way to differentiate the types of noise. This method takes advantage of the visual differences in the noise demonstrated in \Figref{fig:NvY}, and was shown to be capable of detecting both the presence of noise and differentiating the types of noise.

It should be noted that the results presented here were not a parameter scan, and some of the conclusions may depend on the model parameters and especially the duration of the signal. For instance, from \Figref{fig:S_g1_e0.1_mult_dt}, it is evident that the rate of level crossings for $\widetilde{\obs}$ is dependent on $\theta$. In applying the methods from this contribution to experimental data, one should carry out a full Monte Carlo analysis as presented here, using parameters estimated from the PDF and auto-correlation function of the data set.

In this contribution, we have presented an extension of a reference model for intermittent fluctuations in physical systems. Its main application will be for analysis of fluctuation data time series from probe measurements in the scrape-off layer of magnetically confined plasmas, although it is sufficiently general to be applicable elsewhere. This model has no built-in long range correlations, interaction between pulses or other non-stationary effects, meaning it could serve as a useful null hypothesis for investigations of such effects.

\section*{Acknowledgements}

This work was supported with financial subvention from the Research Council of Norway under grant 240510/F20. Discussions with H.~L.~P{\'e}cseli and F.~C.~Geisler are gratefully acknowledged. 

\appendix
\section{List of symbols and results}\label{app:results}

\subsection{Time series}
\begin{tabular}{c | l}
	$\Sn$ & The filtered Poisson process. \\
	$N$ & Normally distributed, uncorrelated noise with vanishing mean and unit standard deviation. \\
	$Y$ & An Ornstein-Uhlenbeck process with vanishing mean and unit standard deviation.\\
	$X$ & A collective term for either $\sigma N$ or $\sigma Y$, where $\sigma$ is a noise intensity parameter.\\
	$\obs$ & Denotes $\Sn + \sigma N$.\\
	$\dyn$ & Denotes $\Sn + \sigma Y$.\\
	$\Snn$ & Denotes $\Sn + X$, it is a collective term for both $\dyn$ and $\obs$.
\end{tabular}

We use the normalization 
\begin{equation}
 \widetilde{\bullet} = \frac{\bullet-\m{\bullet}}{\bullet_\rms}.
\end{equation}

\subsection{Moments and probability density functions}
The PDF of $\Sn$ is a gamma distribution with shape parameter $\gamma$ and scale parameter $\tm{A}$, given by \Eqref{eq:sn_pdf}. Denoting the mean of the process $\Snn$ by $\tm{\Snn}$, its standard deviation by $\Snn_\rms$, its skewness by $S_\Snn$ and its flatness by $F_\Snn$, we have
\begin{align}
  \m{\Snn} &= \gamma \mA,\\
  \Snn_\text{rms}^2 &= (1+\epsilon) \gamma \mA^2,\\
  S_{\Snn} &= \frac{2}{\left( 1+\epsilon \right)^{3/2} \gamma^{1/2}},\\
  F_{\Snn} &= 3+ \frac{6}{\left(1+\epsilon\right)^2 \gamma}.
\end{align}
where $\epsilon = X_\rms^2 / \rSn^2$. Setting $\epsilon = 0$ in the above equation gives the moments of $\Sn$.

With the PDFs of $\Sn$ and $X$ given in \Eqsref{eq:sn_pdf} and \eqref{eq:noise}, and using $\sigma^2 = \epsilon \gamma \mA^2$, we have
\begin{multline}
  \label{pdf_snn}
  P_\Snn(\snn;\gamma,\mA,\epsilon) = 2^{-\gamma/2} (\gamma \epsilon)^{\gamma/2 - 1} \mA^{-1} \exp\left( -\frac{\snn^2}{2 \gamma \epsilon \mA^2} \right) \\ \times \left\{ \frac{(\gamma \epsilon)^{1/2}}{2^{1/2} \Gamma\left( (1+\gamma)/2 \right)} \hyp\left( \frac{\gamma}{2} , \frac{1}{2} ; \frac{1}{2 \epsilon} \left( \frac{\snn}{\gamma^{1/2} \mA}-\gamma^{1/2} \epsilon \right)^2 \right) \right. \\ \left. + \frac{\gamma^{1/2}}{\Gamma\left( \gamma/2 \right)}  \left( \frac{\snn}{\gamma^{1/2} \mA} - \gamma^{1/2} \epsilon \right) \hyp\left( \frac{1+\gamma}{2} , \frac{3}{2} ;\frac{1}{2 \epsilon} \left( \frac{\snn}{\gamma^{1/2} \mA}- \gamma^{1/2} \epsilon\right)^2 \right)  \right\},
\end{multline}
where $\hyp(a,b;z)$ is the confluent hypergeometric function of the first kind, for parameters $a$ and $b$ and argument $z$ \cite{muller}.
Using the normalization in \Eqref{eq:resc_sgnl}, we have that
\begin{equation}
  \nSnn = \left( 1 + \epsilon \right)^{-1/2} \left(  \frac{\Snn}{ \gamma^{1/2} \mA} - \gamma^{1/2} \right).
  \label{eq:def_nsnn}
\end{equation}
We then have
\begin{equation}
  \label{eq:pdf_nsnn_start}
  P_{\nSnn}( \nsnn ) = \left( 1 + \epsilon \right)^{1/2} \gamma^{1/2}  \mA P_{\Snn}\left( \gamma^{1/2} \mA \left[ \left( 1 + \epsilon \right)^{1/2} \nsnn + \gamma^{1/2} \right] \right),
\end{equation}
giving
\begin{multline}
  \label{eq:pdf_nsnn}
  P_{\nSnn}(\nsnn;\gamma,\epsilon) =\left( \frac{\gamma}{2} \right)^{\gamma/2} \epsilon^{\gamma/2 - 1} (1+\epsilon)^{1/2} \exp\left( -\frac{\left[ \left( 1+\epsilon \right)^{1/2} \nsnn + \gamma^{1/2} \right]^2}{2 \epsilon} \right) \\ \times \left\{ \frac{\epsilon^{1/2}}{2^{1/2} \Gamma\left( (1+\gamma)/2 \right)} \hyp\left( \frac{\gamma}{2},\frac{1}{2}; \frac{1}{2 \epsilon} \left[ (1+\epsilon)^{1/2} \nsnn +(1-\epsilon)\gamma^{1/2} \right]^2 \right) \right. \\ \left. + \frac{(1+\epsilon)^{1/2} \nsnn +(1-\epsilon)\gamma^{1/2}}{\Gamma\left( \gamma/2 \right)} \hyp\left( \frac{1+\gamma}{2}, \frac{3}{2} ;\frac{1}{2 \epsilon} \left[ (1+\epsilon)^{1/2} \nsnn +(1-\epsilon)\gamma^{1/2}\right]^2 \right)  \right\}.
\end{multline}
This expression is independent of $\mA$ due to the normalization of $\Snn$. Comparing this distribution to a realization of the process gives $\gamma$ and $\epsilon$ as fit parameters. In the limit of vanishing $\epsilon$, this expression reduces as expected to a Gamma distribution for $\Psi$.

\subsection{Power spectral densities}

The PSD of a random process $\Sn(t)$ is defined as
\begin{equation}
  \psd{\Sn} = \lim_{T \to \infty} \m{\abs{\fourT{\Sn}}^2},
\end{equation}
where
\begin{equation}
  \fourT{\Sn_K} = \frac{1}{\sqrt{T}}\int\limits_{0}^{T} \D t\, \exp(-i \omega t) \Sn(t)
\end{equation}
is the Fourier transform of the random variable over the domain $[0,T]$. Analytical functions which fall rapidly enough to zero [such as the Greens function $G$ given in \Eqref{eq:green}] have the Fourier transform 
\begin{equation}
  \four{G} = \int\limits_{-\infty}^{\infty} \D s\,  G(s) \exp(-i \omega s)
\end{equation}
and the inverse transform
\begin{equation}
G(\tau) = \fourinv{\four{G}} =  \frac{1}{2 \pi} \int\limits_{-\infty}^{\infty} \D \omega \, \exp( i \omega \tau ) \four{G}.
\label{eq:appendix_fourinv}
\end{equation}
We thus have the relevant PSDs,
\begin{align}
\psd{\Sn} &= \rSn^2 \frac{2 \td}{1+\td^2 \omega^2} + 2 \pi \m{\Sn}^2 \delta(\omega),\\
\psd{\dyn} &=  \dyn_\rms^2 \frac{2\td}{1+ \td^2 \omega^2} + 2 \pi \m{\dyn}^2 \delta(\omega),\\
\psd{\obs} &=  \obs_\rms^2 \frac{2 \td}{1+\epsilon} \left[ \frac{1}{1+\td^2 \omega^2} + \frac{\epsilon}{2} \theta  \right] + 2 \pi \m{\obs}^2 \delta(\omega).
\end{align}

\subsection{Auto-correlation functions}
The auto-correlation of a random variable $\Sn$ is the inverse Fourier transform of its PSD, $R_\Sn(\tau) = \fourinv{\psd{\Sn}}$, where the inverse Fourier transform is given by \Eqref{eq:appendix_fourinv}. We thus have
\begin{align}
R_\Sn(\tau) &= \rSn^2 \exp\left( -\frac{\abs{\tau}}{\td} \right) + \m{\Sn}^2,\\
R_\dyn(\tau) &= \dyn_\rms^2 \exp\left( -\frac{\abs{\tau}}{\td} \right) + \m{\dyn}^2,\\
R_\obs(\tau) &= \frac{\obs_\rms^2}{1+\epsilon} \left[ \exp\left( -\frac{\abs{\tau}}{\td} \right) + \epsilon \left( 1-\frac{\abs{\tau}}{\theta \td} \right) \Theta\left( 1-\frac{\abs{\tau}}{\theta \td} \right) \right]+\m{\obs}^2.
\end{align}

\section{Derivation of the power spectral density and auto-correlation function}\label{app:psd}
Results for the autocorrelation function and PSD of a white noise process or Ornstein-Uhlenbeck process are numerous in the literature, see for instance \Refs{uhlenbeck-30,stark-prob,schuss-tasp}. The same results for filtered Poisson processes are also readily available, see e. g. \Refs{rice-44,parzen,lowen-fbpp,garcia-pop-16}. For completeness, we present full derivations in this appendix.

\subsection{Power spectral density of the filtered Poisson process}
To find the PSD of the FPP, we start from \Eqref{eq:shot_noise_green}, and take the Fourier transform
\begin{equation}
  \fourT{\Sn_K} = \frac{1}{\sqrt{T}}\int\limits_{0}^{T} \D t\, \exp(-i \omega t) \Sn_K(t) = \frac{1}{\sqrt{T}}\int\limits_{0}^T \D t\, \exp(-i \omega t)  \int\limits_{-\infty}^{\infty} \D s\, G(s) f_K(t-s).
  \label{eq:def_four}
\end{equation}
where we have exchanged the functions in the convolution given by \Eqref{eq:shot_noise_green}. A change of variables $u(t) = t-s$ gives
\begin{equation}
  \fourT{\Sn_K} = \int\limits_{-\infty}^{\infty} \D s\, G(s) \exp(-i \omega s) \frac{1}{\sqrt{T}} \int\limits_{-s}^{T-s} \D u\, f_K(u) \exp(- i \omega u).
  \label{eq:four_change}
\end{equation}
Note that $G(s)$ is only non-zero for positive $s$ and is negligible after a few $\td$. Moreover, since no pulses arrive for negative times, $f_K(u)=0$ for $u<0$. Assuming $T/\td \gg 1$, we can therefore approximate the limits of the second integral in \Eqref{eq:four_change} as $u \in \left[ 0,T \right]$, and the two integrals become independent. This gives
\begin{equation}
  \fourT{\Sn_K} = \four{G} \fourT{f_K},
  \label{eq:fourierform}
\end{equation}
where
\begin{equation}
  \four{G} = \int\limits_{-\infty}^{\infty} \D s\,  G(s) e^{-i \omega s}.
  \label{eq:four_G}
\end{equation}
The power spectral density (PSD) of the stationary process $\Sn$ is thus
\begin{equation}
  \psd{\Sn} = \lim_{T \to \infty} \m{\abs{\fourT{\Sn_K}}^2} = \m{\abs{\four{G}}^2} \lim_{T \to \infty} \m{\abs{\fourT{f_K}}^2},
  \label{eq:psd_sn_start}
\end{equation}
where $\psd{\Sn}$ is independent of $K$, since the average is over all random variables. The Fourier transform of the Green's functions $\four{G}$ is easily computed as $(i \omega + 1/\td )^{-1}$, giving 
\begin{equation}
  \m{\abs{\four{G}}^2}=\frac{\td^2}{1+\td^2 \omega^2}.
\label{eq:psd_G}
\end{equation}
We also readily find the Fourier transform of the forcing,
\begin{equation}
  \fourT{f_K}=T^{-1/2} \sum_{k=1}^K A_k \exp(-i \omega t_k).
  \label{eq:fourier-f0}
\end{equation}
Multiplying this expression with its complex conjugate and averaging over all random variables gives 
\begin{multline}
  \m{\abs{\fourT{f_K}^2}} = \sum_{K=0}^{\infty} P_K(K;T,\tw) \frac{1}{T}  \sum_{k=1}^K \sum_{l=1}^K \int\limits_0^T \frac{\D t_1}{T} \dots \int\limits_0^T \frac{\D t_K}{T} \\ \times \int\limits_{0}^{\infty} \D A_1 P_A(A_1)\dots\int\limits_{0}^{\infty} \D A_K P_A(A_K) A_k A_l \exp\left( i \omega (t_l-t_k \right)).
  \label{eq:fourier_f}
\end{multline}
In this equation, there are $K$ terms where $k=l$ and $K(K-1)$ terms where $k \neq l$, for which all events are independent. Summing over all these terms, we have
\begin{equation}
  \m{\abs{\fourT{f_K}^2}}= \sum_{K=0}^{\infty}P_K(K;T,\tw) \left[ \frac{K}{T} \m{ {A^2} }+ \frac{K(K-1)}{T^3} \mA^2 \int\limits_0^T \D t \int\limits_0^T \D s \exp(i \omega (t-s)) \right],
  \label{eq:psd_f_0}
\end{equation}
giving
\begin{equation}
  \m{\abs{\fourT{f_K}}^2} = \sum_{K=0}^{\infty}P_K(K;T,\tw) \left[  2 \frac{K}{T} \mA^2 + \frac{2 K (K-1)}{T^3} \mA^2 \frac{1-\cos(\omega T)}{\omega^2} \right]
  \label{eq:psd_f_1}
\end{equation}
where we have used that for an exponentially distributed variable, $\m{ {A^n} } = n! \mA^n$. Thus, averaging over all $K$ and using $\tm{K}=T/\tw$ and $\tm{K(K-1)}=T^2/\tw^2$ gives
\begin{equation}
  \m{\abs{\fourT{f_K}}^2} = \frac{2}{\tw} \mA^2 + \frac{2}{\tw^2} \mA^2 \frac{1-\cos(\omega T)}{ T \omega^2}.
  \label{eq:psd_f_2}
\end{equation}
The second term in this equation resembles a Dirac delta function in the limit $T \to \infty$. With the appropriate normalization $\int_{-\infty}^{\infty}\D \omega \, \delta(\omega) = 1$ this gives
\begin{equation}
  \lim_{T \to \infty} \m{\abs{\fourT{f_K}}^2} = \frac{2}{\tw} \mA^2 + \frac{2 \pi}{\tw^2} \mA^2 \delta(\omega),
  \label{eq:psd_f}
\end{equation}
which together with \Eqref{eq:psd_G} gives the PSD of the FPP as
\begin{align}
  \psd{\Sn} &= 2 \gamma \mA^2 \frac{\td}{1+ \td^2 \omega^2} + 2 \pi \gamma^2 \mA^2 \delta(\omega) \nonumber\\
  &= \rSn^2 \frac{2 \td}{1+\td^2 \omega^2} + 2 \pi \m{\Sn}^2 \delta(\omega).
  \label{eq:psd_sn_fin}
\end{align}

\subsection{Power spectral density and auto-correlation function of the noise processes}\label{app:psd_noise}
Since $Y(t)$ is constructed as a convolution in the same way as the FPP, we have an analogue of \Eqref{eq:psd_sn_start} for this process:
\begin{equation}
  \psd{Y} = \m{\abs{\four{G}}^2} \lim_{T \to \infty} \m{\abs{\fourT{\D W}}^2}.
  \label{eq:psd_Y_start}
\end{equation}
With \Eqref{eq:psd_G} and the relation
\begin{equation}
  \lim_{T \to \infty} \m{\abs{\fourT{\D W}}^2} = \lim_{T \to \infty} \frac{2}{\td T} \m{ \int\limits_0^T \D W(t) \exp(-i \omega t) \int \limits_0^T \D W(s) \exp(i \omega s) } = \frac{2}{\td},
  \label{eq:psd-y-mid}
\end{equation}
we have  the power spectral density for the Ornstein-Uhlenbeck process,
\begin{equation}
  \psd{Y} = \frac{ 2 \td}{1+\td^2 \omega^2}.
  \label{eq:psd_Y}
\end{equation}
The power spectral density of $Y(t)$ is therefore Lorentzian with the same parameter as for the PSD of $\Sn(t)$, implying that the auto-correlation function of $Y(t)$ is also an exponentially decaying function with the same rate as the auto-correlation function of $\Sn(t)$:
\begin{equation}
  R_Y(\tau) = \exp\left(-\frac{\abs{\tau}}{\td}\right).
  \label{eq:R_Y}
\end{equation}

For observational noise we require that $N(t)$ is a unit-less variable. One way of realizing such a process is by using integrated increments of the Wiener process:
\begin{equation}
  N(t) = \frac{1}{\dt^{1/2}} \int\limits_t^{t+\dt} \D W(s),
  \label{eq:N_def}
\end{equation}
where $\dt$ is the sampling time (in this case, each sample $N[n] = N(n \dt)$, $n = 0,1,2,\dots$ is normally and independently, identically distributed with zero mean and unit standard deviation). In this case, the most direct route to the power spectral density is via the auto-correlation function. We find that
\begin{equation}
  R_N(\tau) = \left(1-\frac{\abs{\tau}}{\dt}\right) \Theta\left(\dt- \abs{\tau} \right),
  \label{eq:acorr_N}
\end{equation}
giving
\begin{equation}
  \psd{N} = \int\limits_{-\infty}^{\infty} \D \tau\, \exp\left( -i \omega \tau \right) R_N(\tau) = 2 \frac{1-\cos\left( \dt \omega \right)}{\dt \omega^2}.
  \label{eq:psd_N_0}
\end{equation}
Using the normalized sampling time $\theta$ from \Eqref{eq:norm-time-step}, this expression can be written as
\begin{equation}
  \psd{N} = 2 \frac{\td}{\theta} \frac{1- \cos\left( \theta \, \td \omega \right)}{\td^2 \omega^2}.
  \label{eq:psd_N}
\end{equation}
For small $\theta$, the cosine function can be expanded around 0, and we have
\begin{equation}
  \lim_{\theta \to 0} \frac{\psd{N}}{\td \theta} = 1.
  \label{eq:psd-N-theta-0}
\end{equation}
In this limit, $N(t)$ approaches white noise, which has a flat power spectrum. Since we have demanded $N_\rms = \int_{-\infty}^{\infty} \D \omega \psd{N}^2 = 1$, the higher resolution granted by $\theta \to 0$ means the total power of $N$ is divided among a greater number of frequencies, reducing the power per frequency.

Since $\dt$ is the smallest time we will observe, it makes more sense to use the i.i.d. random sequence $N[n]$ and to translate the auto-correlation to discrete time:
\begin{equation}
  R_N[n] = \begin{cases} 1 & \text{, } n = 0 \\
 	0 & \text{, } n \geq 1 \\
      \end{cases}.
  \label{eq:acorr_N_discrete}
\end{equation}
The discrete Fourier transform corresponding to $\four{\bullet}$ is then
\begin{equation}
  \psd{N} = \dt \sum\limits_{n = -\infty}^{\infty} \exp\left( - i \omega n \right) R_N[n] = \td \theta.
  \label{eq:psd_N_discrete}
\end{equation}
Intuitively, this corresponds to the case where we don't see the effects of $\theta$, that is the regime $\theta \ll 1$. Since this is also the spectral density we will observe for a realization of the process $N(t)$, we will use \Eqref{eq:psd_N_discrete} when discussing the power spectral density of $\Omega(t)$.

\bibliography{sources}
\bibliographystyle{apsrev4-1} 
\end{document}